\documentclass[11pt,a4paper]{article}
\usepackage{jcappub}
\newcommand{\be}{\begin{eqnarray}}
\newcommand{\ee}{\end{eqnarray}}
\newcommand{\simgt}{\lower.5ex\hbox{$\; \buildrel > \over \sim \;$}}
\newcommand{\simlt}{\lower.5ex\hbox{$\; \buildrel < \over \sim \;$}}
\newcommand{\rmd}{{\rm d}}
\newcommand{\fnl}{f_{\rm nl}}
\newcommand{\tnl}{\tau_{\rm nl}}

\newcommand{\gnl}{g_{\rm nl}}
\newcommand{\hnl}{h_{\rm nl}}
\newcommand{\inl}{i_{\rm nl}}

\newcommand{\chio}{{\chi_1}}
\newcommand{\chit}{{\chi_2}}
\newcommand{\chii}{{\chi_i}}
\newcommand{\chiol}{{\chi_{1,\ell}}}
\newcommand{\chitl}{{\chi_{2,\ell}}}
\newcommand{\chiil}{{\chi_{i,\ell}}}

\newcommand{\chios}{{\chi_{1,s}}}
\newcommand{\chits}{{\chi_{2,s}}}
\newcommand{\chiis}{{\chi_{i,s}}}
\newcommand{\vx}{{\bf x}}
\newcommand{\vy}{{\bf y}}
\newcommand{\vk}{{\bf k}}
\newcommand{\vp}{{\bf p}}
\newcommand{\vq}{{\bf q}}
\newcommand{\vr}{{\bf r}}

\newcommand{\perm}[1]{(#1\,{\rm perms.})}
\newcommand{\deltac}{\delta_{\rm c}}

\newcommand{\Mk}{\mathcal{M}(k)}
\newcommand{\Mr}{\mathcal{M}_R}

\newcommand{\dk}{\delta_\vk}
\newcommand{\zk}{\zeta_\vk}
\newcommand{\zx}{\zeta(\vx)}

\title{Scale Dependence of the Halo Bias in General Local-Type Non-Gaussian Models I: Analytical Predictions and Consistency Relations}
\author{Takahiro Nishimichi}
\affiliation{Kavli Institute for the Physics and Mathematics of the Universe, 
The University of Tokyo,\\
5-1-5 Kashiwanoha, Kashiwa, Chiba 277-8568, Japan}
\emailAdd{takahiro.nishimichi@ipmu.jp}

\abstract{
We investigate the clustering of halos in cosmological models starting 
with general local-type non-Gaussian primordial fluctuations. 
We employ multiple Gaussian fields and add local-type non-Gaussian corrections at arbitrary order
to cover a class of models described by frequently-discussed $\fnl$, $\gnl$ and $\tnl$ parameterization.
We derive a general formula for the halo power spectrum based on the peak-background split formalism. 
The resultant spectrum is characterized by only two parameters responsible for the scale-dependent bias at large scale
arising from the primordial non-Gaussianities in addition to the Gaussian bias factor.
We introduce a new inequality for testing non-Gaussianities originating from multi fields, 
which is directly accessible from the observed power spectrum. 
We show that this inequality is a generalization of the Suyama-Yamaguchi inequality 
between $\fnl$ and $\tnl$ to the primordial non-Gaussianities at arbitrary order.
We also show that the amplitude of the scale-dependent bias is useful
to distinguish the simplest quadratic non-Gaussianities (i.e., $\fnl$-type) from higher-order ones ($\gnl$ and higher),
if one measures it from multiple species of galaxies or clusters of galaxies.
We discuss the validity and limitations of our analytic results by comparison with numerical simulations 
in an accompanying paper.
}

\keywords{large scale structure, primordial non-Gaussianity, numerical simulation}

\arxivnumber{1204.3490}

\notoc
\begin{document}
\maketitle
\flushbottom

%===========================================================
\section{Introduction}
%===========================================================

%Forecasts:
%\cite{Carbone08,Verde09,Verde10,Fedeli11,Giannantonio11}

The large-scale structure (LSS) of the universe as well as the temperature fluctuations of 
the cosmic microwave background (CMB) radiation are well understood in the standard framework of 
cosmological model starting from tiny almost Gaussian fluctuations.
However, the generation mechanism of the primordial fluctuations, which seed both the CMB and the LSS, 
is yet to be understood.
Because we usually consider that the primordial perturbations are created during inflation, their statistical 
properties provide us valuable information about the inflationary physics.
Although we have no clear evidence of the violation of the Gaussian assumption of our initial condition to date,
the signature of the primordial non-Gaussianities, if they are detected, can be a smoking gun to test various 
models of the primordial fluctuations in the coming era of precision cosmology.

Some proposed models of the early universe predict that 
large local-type non-Gaussianities are produced when nonlinear dynamics is 
important on super-horizon scales \cite{Bartolo04}.
In these models, the Bardeen's curvature potential $\Phi$ in the matter dominant era 
is written as a local function of a Gaussian field $\phi$, and is usually expanded into a Taylor series as
%--------------
\be
\Phi(\vx) = \phi(\vx) + \fnl \left[\phi^2(\vx) - \langle\phi^2\rangle\right] + \gnl\,\phi^3(\vx) + \hnl\left[\phi^4(\vx)-\langle
\phi^4\rangle\right] + \inl\,\phi^5(\vx)+\cdots.
\label{eq:Bardeen}
\ee
%--------------
The coefficient $\fnl$ in the quadratic term is one of a key parameter to distinguish the 
models\footnote{Note that we adopt the normalization of $\fnl$ widely used in studies of the CMB, instead of the convention
sometimes employed in the LSS works, where this parameter is also defined by Eq.~(\ref{eq:Bardeen}), but the normalization
of the curvature potential is done at the present by linear extrapolation.} 
\cite{Komatsu01}.
This parameter determines the amplitude of the primordial bispectrum, the lowest-order statistic that captures
the deviation from Gaussian statistics. In this model, the bispectrum of the curvature perturbation $\zeta \equiv (3/5)\Phi$
is written as
%----------
\be
B_\zeta(k_1,k_2,k_3) &=& \frac65\fnl\left[P_\zeta(k_1)P_\zeta(k_2)+\perm{2}\right],\label{eq:Bfnl}
\ee
%----------
where $P_\zeta$ denotes the power spectrum of $\zeta$ and 
$\perm{2}$ stands for two more terms given as permutations of wavenumbers, $k_1$, $k_2$ and $k_3$.
The current constraints on the parameter $\fnl$ from the CMB experiments are very tight, and any deviation from Gaussianity 
of order $\simgt 0.1\%$ level in the curvature perturbations has already been ruled out (e.g., \cite{WMAP7}).
It is expected that the constraint will be much tighter after some on-going/upcoming large observational 
projects such as Planck \cite{Planck}.

However, the value of $\fnl$ does not determine all the statistical properties of the initial condition of our universe, 
even if one focuses only on the local-type non-Gaussian primordial fluctuations.
The next order statistic, the trispectrum, is a natural extension of the discussion, and it indeed
gives us more hints about the inflationary physics.
The cubic-order correction to the curvature potential is a natural source of this statistic 
\cite{Okamoto02,Enqvist05} ($\gnl$-type; see Eq.~\ref{eq:Bardeen}).
This gives the trispectrum of the form
%----------
\be
T_\zeta(k_1,k_2,k_3,k_4) = \frac{54}{25}\gnl\left[P_\zeta(k_1)P_\zeta(k_2)P_\zeta(k_3)+\perm{3}\right].
\label{eq:Tgnl}
\ee
%----------
The trispectrum can also be generated through the quadratic corrections \cite{Boubekeur06}.
This type of trispectrum has a different scale dependence from Eq.~(\ref{eq:Tgnl}) and is written as
%----------
\be
&&T_\zeta(k_1,k_2,k_3,k_4) = \tnl\left[P_\zeta(k_1)P_\zeta(k_2)P_\zeta(\left|\vk_1+\vk_3\right|)+\perm{11}\right],
\label{eq:Ttnl}
\ee
%----------
where the parameter $\tnl$ controls the amplitude of the spectrum. In single-field models described by
Eq.~(\ref{eq:Bardeen}), $\tnl$ can be expressed by $\fnl$ as $\tnl = (36/25)\fnl^2$.
In general, however, when one employs two or more Gaussian components in the primordial fluctuations, 
$\tnl$ can be larger, and the following inequality is shown to be always valid 
(\cite{Suyama08}; Suyama-Yamaguchi inequality):
%----------
\be
\tnl \geq \frac{36}{25}\fnl^2,
\ee
%----------
where $\fnl$ is defined not by Eq.~(\ref{eq:Bardeen}) but by Eq.~(\ref{eq:Bfnl}) as the amplitude of the 
bispectrum of the curvature perturbations.
See \cite{Suyama10,Sugiyama11,Smith11c} for further discussions on the loop corrections and
the universality of this inequality, and \cite{Smidt10} for an observational attempt to test it from existing observations.
Thanks to this inequality, one can distinguish multi-field inflationary models from single-field ones by 
measuring $\fnl$ and $\tnl$ simultaneously.
The statistical properties of the primordial perturbations are uniquely determined by three parameters,
$\fnl$, $\gnl$ and $\tnl$, up to the order of trispectrum as long as one is interested 
in local-type non-Gaussianities\footnote{Note, however, that these parameters can be scale dependent in general (see, e.g., 
\cite{Byrnes10a,Byrnes10b}).}, and a number of models can be sorted by these parameters \cite{Suyama10}. 

The LSS that we observe through the clustering of galaxies or clusters of galaxies is a highly non-Gaussian field
due to the nonlinear gravitational evolution (e.g., \cite{Bernardeau02} for a review), 
redshift-space distortions caused by the peculiar velocity field of the galaxies \cite{Kaiser87} and 
the galaxy bias with respect to the underlying matter density field \cite{Kaiser84}.
These acquired non-Gaussianities are thought to prevent us to measure only the primordial non-Gaussianity
from the LSS unlike the CMB, whose temperature fluctuations are still in the linear stage.
It is thus not straightforward to extract the non-Gaussian signature of purely primordial origin from the LSS, and
a number of studies have been done to model the gravitationally generated non-Gaussianities
\cite{Scoccimarro00,Verde00,Verde01,Scoccimarro04b,Sefusatti07}.
In these studies, the authors mainly focus on the galaxy bispectrum, simply because it is the lowest-order 
statistic that captures the non-Gaussian signature (but see \cite{Verde01} for a discussion on the trispectrum). 

However, one observable feature in the LSS, which was realized only recently, attracts a great attention \cite{Dalal08}.
That is the {\it scale-dependent bias} in the halo power spectrum on large scales (i.e., $\simgt$Gpc).
%In short, this scale dependence comes from the mode mixing between large and small 
%scale due to the non-Gaussianities in the initial condition. 
%The small scale modes which explain the collapse of halos at later time are modulated by modes 
%on larger scales, and this results in the clustering of halos with a strong feature at large scale. 
This is an interesting example where the bias, which usually prevents us from extracting cosmological
information, opens a new window to see the primordial bispectrum but in the power spectrum.
This new feature has been extensively studied both analytically 
\cite{Matarrese08,Afshordi08,Taruya08b,McDonald08,Giannantonio10,Valageas10,Desjacques11a,Desjacques11b} and numerically \cite{Grossi09,Pillepich10,Nishimichi10}, and gives stringent constraints on
$\fnl$ \cite{Slosar08,Xia10a,Xia10b}, which are already competitive to those from the CMB.
Also, the scale-dependent bias affects the halo bispectrum similarly, and thus there is a possibility to
put even stronger constraints on $\fnl$ from the LSS by combining the power spectrum with the bispectrum
\cite{Jeong09,Sefusatti09,Nishimichi10,Baldauf11,Sefusatti11}.

A similar scale-dependent halo bias has been reported in the presence of the primordial trispectrum
(see \cite{Desjacques10a,Smith11b,Chongchitnan11,Gong11} for $\gnl$-type and 
\cite{Tseliakhovich10,Smith11a,Gong11} for $\tnl$-type). 
Furthermore, the scale dependence in the halo bias is found to be present in other models of the primordial
non-Gaussianities, such as the scale-dependent $\fnl$ model 
\cite{Becker11,Shandera11}, ungaussiton model \cite{Yokoyama11b}, 
and non-local non-Gaussian models \cite{Schmidt10,Wagner11,Scoccimarro11}.
Thus the interpretation is not straightforward even when one detects a clear evidence of the scale-dependent bias 
from observed clustering of galaxies or clusters of galaxies.  

Our aim is to develop a statistical methodology to distinguish these models.
Moreover, we would like to generalize the argument as model-independent as possible.
In this paper, as a first attempt, we focus on the scale-independent local-type non-Gaussianities and discuss 
the halo clustering. In particular, we employ multiple Gaussian fields that seed the curvature perturbations in order to
control $\fnl$ and $\tnl$ independently. We also keep all the higher-order corrections in the curvature perturbations
such as $\gnl$ and $\hnl$ to see how generic our result is.
Our model is a simple generalization of the $\fnl$, $\gnl$ and $\tnl$ parameterization.
We will show that this class of primordial non-Gaussianities result in a universal formula for the halo power spectrum.
We introduce a new parameterization of the non-Gaussian 
signature in the scale-dependent bias and propose two tests to distinguish different models based on the analytic results.
The validity and the limitations of our analytical results are tested by confronting with a large set of cosmological
$N$-body simulations, and will be presented in a separate paper (hereafter paper II).

This paper is organized as follows. We first introduce our model of the primordial non-Gaussianities 
and discuss the basic statistical properties
in Sec.~\ref{sec:model}. We then compute the halo power spectrum following the peak-background
split argument in Sec.~\ref{sec:analytic}. We discuss the two relevant parameters for the feature in the 
scale-dependent bias and show how they are useful to distinguish different models in Sec.~\ref{sec:obs}.
We give a brief summary of the paper in Sec.~\ref{sec:summary}.
Throughout the paper, we adopt the best-fit flat $\Lambda$CDM cosmology to the seven-year observations of WMAP
\cite{WMAP7} and adopt $A_s=2.43\times10^{-9}$, $n_s=0.963$ for the amplitude and tilt of the power spectrum
of the curvature perturbations normalized at wavenumber $k_0=0.002\,$Mpc$^{-1}$, $\Omega_{\rm m}=0.265$,
$\Omega_{\rm b}=0.0448$ for the density parameters of total matter and baryon and $h=0.71$ for the Hubble parameter.
We compute the transfer function from the curvature perturbation to the matter density by {\tt CAMB} \cite{CAMB}
using these parameters.

%===========================================================
\section{The model}
\label{sec:model}
%===========================================================
In this section, we describe our model of the non-Gaussian curvature perturbations and
show their basic statistical properties. This section gives useful quantities to derive
the clustering of halos in later sections, such as the cumulants of the density field. 
We first introduce our parameterization for the curvature perturbations in Sec.~\ref{subsec:local}, 
and compute the polyspectra (Sec.~\ref{subsec:spectrum}) and cumulants (Sec.~\ref{subsec:cumulants}).
%===========================================================
\subsection{Local-type non-Gaussianities with two fields}
\label{subsec:local}
%===========================================================
In this subsection, we introduce our models of the non-Gaussian primordial perturbations.
Throughout the paper, we restrict our attention to the adiabatic density fluctuations, 
whose statistical properties are solely transferred from the primordial curvature perturbations:
%----------
\be
\dk(z) = \mathcal{M}(k,z) \zk,
\label{eq:transfer}
\ee
%----------
where $\mathcal{M}(k,z)$ denotes the matter transfer function, and is normalized such that 
$\dk(z)$ gives the linear overdensity field at redshift $z$. We hereafter omit the redshift dependence,
but it always comes through $\mathcal{M}(k,z)$.
Then, we consider the local-type non-Gaussianities
in the curvature perturbations originating from two Gaussian fields. We expand the curvature perturbations into
a Taylor series:
%----------
\be
\zx = \chio(\vx)+\chit(\vx)+\sum_{i,j}c_{i,j}\left[\chi_1^i(\vx)\chi_2^j(\vx) - \langle\chi_1^i\chi_2^j\rangle\right],
\label{eq:curvature}
\ee
%----------
where $\chio$ and $\chit$ are two statistically independent auxiliary Gaussian fields, $\langle\chio\chit\rangle=0$.
In the above, the indices $i$ and $j$ in the summation run over positive integers with $i+j\geq2$.
We assume that the linear terms in Eq.~(\ref{eq:curvature}) are dominant contributions and 
the higher-order terms give little corrections to the total curvature perturbations.

After the main discussion starting with the two-field model in Eq.~(\ref{eq:curvature}) we
will show how the result can be generalized to multi-field models with $N\geq2$ independent
Gaussian fields. Our final results for the halo power spectrum do not depend on the number of
fields as will be shown in Sec.~\ref{subsec:halo}.

%===========================================================
\subsection{Polyspectra}
\label{subsec:spectrum}
%===========================================================
Let us first compute the polyspectra based on Eq.~(\ref{eq:curvature}).
We focus on models in which the two fields, $\chio$ and $\chit$, have the same spectral index,
and parameterize their power spectra as
%----------
\be
P_\chio(k) &=& (1-\alpha)\,P_\zeta(k),\\
P_\chit(k) &=& \alpha\,P_\zeta(k),\\
P_\zeta(k) &\equiv& \frac{2\pi^2}{k^3}A_{\rm s}\left(\frac{k}{k_0}\right)^{n_{\rm s}-1},
\label{eq:Pzeta}
\ee
%----------
Note that Eq.~(\ref{eq:curvature}) reduces to the single-field model (\ref{eq:Bardeen}) when $\alpha=0$.
We can generalize the model (\ref{eq:curvature}) into scale-dependent non-Gaussianities by setting 
different tilts for the two fields, and these models will be discussed elsewhere.

We then consider the higher-order polyspectra of $\zeta$ at the tree-level. 
The model (\ref{eq:curvature}) reproduces the bispectrum of the curvature perturbations 
expressed by Eq.~(\ref{eq:Bfnl}), with the amplitude parameter $\fnl$ being the following form:
%----------
\be
\fnl &=& \frac53\left[(1-\alpha)^2c_{2,0}+\alpha(1-\alpha)c_{1,1}+\alpha^2c_{0,2}\right].
\label{eq:fnl}
\ee
%----------
Similarly, the model gives the trispectrum written as the sum of Eqs.~(\ref{eq:Tgnl}) and (\ref{eq:Ttnl}), 
and the relevant parameters read
%----------
\be
\gnl &=& \frac{25}9\left[(1-\alpha)^3c_{3,0} + \alpha(1-\alpha)^2c_{2,1} + \alpha^2(1-\alpha)c_{1,2} + \alpha^3c_{0,3}\right],\label{eq:gnl}\\
\tnl &=& 4(1-\alpha)^3c_{2,0}^2+\alpha(1-\alpha)c_{1,1}^2+4\alpha^3c_{0,2}^2 + 
4\alpha(1-\alpha)c_{1,1}\left[(1-\alpha)c_{2,0}+\alpha c_{0,2}\right].\label{eq:tnl}
\ee
%----------
The trispectrum has two terms, one comes from the cubic coupling, $c_{i,j}$ with $i+j=3$, and is characterized by $\gnl$,
while the other originates from $c_{i,j}$ with $i+j=2$ (i.e., the $\tnl$ term).
If one specifies the values of $\fnl$, $\gnl$ and $\tnl$, the primordial trispectrum as well as the bispectrum
are uniquely determined. 

In this paper, we further discuss the primordial non-Gaussianities beyond the trispectrum level to confirm the generality
of the conclusion. In order to do so, we compute the tetraspectrum (the fifth order spectrum) of the curvature perturbations.
In addition to the contribution from $\hnl$ in Eq.~(\ref{eq:Bardeen}), we find two more terms arising 
from lower-order non-Gaussianities (i.e., $\fnl$ and $\gnl$). They are shown in Appendix~\ref{app:quartic}.
As a specific example, we investigate the effect of quartic-order non-Gaussianity characterized by 
$\hnl$, and the results of $N$-body simulations will also be presented in paper II.

%===========================================================
\subsection{Cumulants}
\label{subsec:cumulants}
%===========================================================

The cumulants of the density field are basic quantities useful in analytically estimating the halo number density 
(see Sec.~\ref{subsec:halo} below).
In this subsection, we compute the low-order cumulants and give their fitting formulae.
%(note that the exact computation includes
%multi-dimensional integrals, and thus is computationally expensive).

We begin by defining the smoothed density field in Fourier space:
%----------
\be
\delta_{R,\vk} \equiv \widetilde{W}_R(k)\delta_\vk,
\label{eq:deltaR}
\ee
%----------
where $\widetilde{W}_R$ is the Fourier transform of the window function at scale $R$, and we adopt the top hat function.
We then define its $n$-th order cumulant (or, the connected moment) as
%----------
\be
\bar{\kappa}^{(n)}_R \equiv \langle\delta_R^n\rangle_{\rm c}.\label{eq:cumulant}
\ee
%----------
We introduce the windowed transfer function, $\Mr(k) \equiv \widetilde{W}_R(k)\mathcal{M}(k)$, which directly connects
the smoothed density contrast to the curvature perturbation:
%----------
\be
\delta_{R,\vk} = \Mr(k)\zk.
\label{eq:transfer_R}
\ee
%----------
Using Gaussianity of $\chio$ and $\chit$, the leading order non-vanishing cumulants are computed as
%----------
\be
\bar\kappa^{(n)}_R = \sum_{i_1,\ldots,i_n,j_1,\ldots,j_n} \left[\prod_{m=1}^n c_{i_m,j_m}\right] 
\left\langle \prod_{m=1}^n\left[\Mr * \left(\chi_1^{i_m}\chi_2^{j_m} - \langle\chi_1^{i_m}\chi_2^{j_m}\rangle
\right)\right]\right\rangle_{\rm c},
\label{eq:cumulants}
\ee
%----------
where the product $*$ represents a convolution with the windowed transfer function:
%----------
\be
(\Mr*A)(\vx) &=& \int\rmd^3\vy \Mr(\vx-\vy)A(\vy),\label{eq:conv_Mr},
\ee
%----------
with $\Mr(\vx)$ being the Fourier counterpart of $\Mr(k)$.
%%----------
%\be
%\Mr(\vx) &\equiv& \int\frac{\rmd^3\vk}{(2\pi)^3}\Mr(k)e^{i\vk\cdot\vx}.
%\ee
%%----------
In Eq.~(\ref{eq:cumulants}), the summation runs for $2n$ non-negative integers, $i_m$ and $j_m$, which satisfy
%----------
\be
&&i_m + j_m \geq 1,\qquad {\rm for}\,\,\,m=1, 2, \ldots, n,\\
&&\sum_{m=1}^n (i_m + j_m) = 2 (n-1).
\ee
%----------
They can be explicitly rewritten by integrals of the polyspectra of $\zeta$, already computed
in the previous subsection, after multiplying the window function. 
For example, the first three cumulants, the variance $\bar{\kappa}^{(2)}_R$, the skewness $\bar{\kappa}^{(3)}_R$ 
and the kurtosis $\bar{\kappa}^{(4)}_R$, 
are given by
%----------
\be
\bar{\kappa}^{(2)}_R &=& \int\frac{\rmd^3\vp}{(2\pi)^3}\Mr(p)^2 P_\zeta(p) \equiv \bar\sigma^2_R,\\
\bar{\kappa}^{(3)}_R &=& \int\frac{\rmd^3\vp\rmd^3\vq}{(2\pi)^6}
\Mr(p)\Mr(q)\Mr(|\vp+\vq|)B_\zeta(p,q,|\vp+\vq|) \equiv \fnl\,\hat{\kappa}^{(3)}_{R,\fnl},\\
\bar{\kappa}^{(4)}_R &=& \int\frac{\rmd^3\vp\rmd^3\vq\rmd^3\vr}{(2\pi)^9}
\Mr(p)\Mr(q)\Mr(r)\Mr(|\vp+\vq+\vr|)T_\zeta(\vp,\vq,\vr,-\vp-\vq-\vr),\nonumber\\
&\equiv&  \gnl\,\hat{\kappa}^{(4)}_{R,\gnl} + \tnl\,\hat{\kappa}^{(4)}_{R,\tnl},
\ee
%----------
where $\hat{\kappa}^{(4)}_{R,\gnl}$ and $\hat{\kappa}^{(4)}_{R,\tnl}$ originate from the trispectrum
in Eq.~(\ref{eq:Tgnl}) and (\ref{eq:Ttnl}), respectively.
We find that the following fitting formulae for the skewness and kurtosis accurately reproduce the numerical 
integrals of the above equations for our fiducial cosmology at $z=0$:
%----------
\be
\hat{\kappa}^{(3)}_{R,\fnl} \simeq 3.20\times10^{-4}\,\bar\sigma^{3.141}_R, \label{eq:fit_fnl}\\
\hat{\kappa}^{(4)}_{R,\gnl} \simeq 5.81\times10^{-8}\,\bar\sigma^{4.297}_R,\label{eq:fit_gnl}\\
\hat{\kappa}^{(4)}_{R,\tnl} \simeq 2.06\times10^{-7}\,\bar\sigma^{4.288}_R.\label{eq:fit_tnl}
\ee
%----------
See also \cite{Chongchitnan10} and \cite{LoVerde11} for similar fitting formulae.
We also derive analogous expression for the fifth-order cumulant, $\bar{\kappa}^{(5)}_R$.
It can be found in Appendix~\ref{app:quartic}. 

%===========================================================
\section{Analytical prediction of the halo clustering}
\label{sec:analytic}
%===========================================================
So far, we have
%not said anything about the formation or the clustering of halos, but 
focused on the basic properties 
of the underlying density field as well as the curvature perturbations.
Now, let us consider a simple model of the halos in the presence of general local-type non-Gaussianities 
described by Eq.~(\ref{eq:curvature}).

We follow the peak-background split formalism, and generalize it to our model, Eq.~(\ref{eq:curvature}) in
Sec.~\ref{subsec:PBS}.
There, we define the conditional cumulants and show how these cumulants are affected by primordial non-Gaussianities.
After that, we discuss
% the halo mass function and describe a simple model in Sec.~\ref{subsec:halo}.
%The idea of the halo mass function is extended to a conditional one to finally give 
the clustering properties of halos based on the calculation of the conditional number density of them
in Sec~\ref{subsec:halo}.

%===========================================================
\subsection{The peak-background split in the presence of non-Gaussianities}
\label{subsec:PBS}
%===========================================================
%===========================================================
\subsubsection{long and short mode decomposition}
\label{subsubsec:longshort}
%===========================================================

The peak-background split formalism had been originally introduced to model the biased clustering of
peaks, halos, or galaxies which are associated with the initial peak regions in the Lagrangian space
\cite{Bardeen86,Cole89,Mo96,Catelan98}.
This formalism gives a simple framework to compute the bias of the collapsed objects with mass $M$ by 
considering the {\it conditional} probability that a density contrast with a mode of scale $R$ exceeds a 
critical density $\deltac$ under the background density field modulated by long-wavelength modes.
Here, the mass and the scale are related as
%----------
\be
R \equiv \left(\frac{3M}{4\pi\bar\rho}\right)^{1/3},\label{eq:mass2rad}
\ee
%----------
where $\bar\rho$ denotes the cosmic mean density.
Effectively, the presence of long modes can be interpreted as a local shift in the threshold density:
%----------
\be
\deltac \to \deltac(\vq) \equiv \deltac - \delta_\ell(\vq),\label{eq:deltac_effective}
\ee
%----------
where the coordinate $\vq$ indicates that we are working in the Lagrangian space.
We adopt $\deltac=1.686$ following the result of the spherical collapse model, 
and leave the discussion on its more accurate modeling to paper II.
Using Eq.~(\ref{eq:deltac_effective}), one can relate the long mode to the halo number density at that position 
with a help of the halo mass function.

%In case of Gaussian initial conditions, we start with decomposing the linear density field into a sum of long and short 
%wavelength contributions:
%%----------
%\be
%\delta(\vx) = \delta_\ell(\vx) + \delta_s(\vx).
%\ee
%%----------
%Note that we adopt the Lagrangian picture and the above decomposition is done in the Lagrangian space.
%Here the two modes ``$\ell$" and ``{\it s}" are statistically independent since $\delta$ is a Gaussian field.
%We consider that the short mode 
%``$s$" is responsible for the formation of halos, and the mass of halos $M$ is 
%related to the length scale $R$ by
%%----------
%\be
%R \equiv \left(\frac{3M}{4\pi\bar\rho}\right)^{1/3},\label{eq:mass2rad}
%\ee
%%----------
%where $\bar\rho$ denote the cosmic mean density.
%Thus variables with a subscript ``$s$" depends on a scale $R$ (or mass $M$) implicitly.
%The long mode, on the other hand, does not directly contribute to the halo formation, but adds a small 
%modulation to the background density.
%This modulation can be reinterpret as a change in the effective threshold 
%for the collapse:
%%----------
%\be
%\deltac \to \deltac(\vx) \equiv \deltac - \delta_\ell(\vx),\label{eq:deltac_effective}
%\ee
%%----------
%where $\deltac$ denotes the linear critical density for the formation of halos.
%We adopt $\deltac=1.686$ following the result of the spherical collapse model, 
%and leave the discussion on its more accurate modeling to paper II.

In the presence of primordial non-Gaussianities, Eq.~(\ref{eq:deltac_effective}) is not the only effect
from the long modes, since Fourier modes are not independent of each other from the beginning.
We will show how this mode coupling results in the statistical properties of the short modes that are responsible for
the formation of halos.

We start with decomposition of the two auxiliary Gaussian fields into sums of the long and short wavelength 
contributions:
%----------
\be
\chii(\vq) = \chiil(\vq) + \chiis(\vq),\qquad i=1,\,2.
\ee
%----------
Then, keeping only the linear modulation from $\chiil$, the short mode for the density fluctuation can be computed as
%----------
\be
\delta_s(\vq) &=& \sum_{i,j} \left[c_{i,j}+\Delta c_{i,j}(\vq)\right](\Mr*\chios^i\chits^j)(\vq),
\label{eq:deltas}\\
\Delta c_{i,j}(\vq) &\equiv& (i+1)c_{i+1,j}\,\chiol(\vq)+(j+1)c_{i,j+1}\,\chitl(\vq),
\label{eq:delta_cij}
\ee
%----------
where $c_{1,0} = c_{0,1} = 1$ and the convolution $*$ is defined by Eq.~(\ref{eq:conv_Mr}).
Interestingly, one can see that a $(n+1)$th-order coupling constant in $\zeta$ (i.e., $c_{i,j}$ with $i+j=n+1$) 
modifies $\delta_s$ at the $n$th order. For instance, in case of quadratic non-Gaussianities characterized by $\fnl$,
the short mode is affected by the long mode at the linear order (i.e., the amplitude of the power spectrum is modulated), 
and this gives the scale-dependent bias \cite{Slosar08}.
We will see how the statistical properties of the conditional density field (\ref{eq:deltas}) result in the characteristic
feature in the halo clustering shortly.

%===========================================================
\subsubsection{conditional cumulants}
\label{subsubsec:localcumulants}
%===========================================================
In Sec.~\ref{subsec:cumulants}, we have computed the cumulants of the density field in the presence of 
primordial non-Gaussianities.
The cumulants, denoted by $\bar{\kappa}^{(n)}_R$, are given as the volume averages of the products of the density
contrast over the entire universe.
One of the key idea in the peak-background split formalism is to consider the {\it conditional} cumulants of the short mode 
at an arbitrary position $\vq$ in the presence of the long modes. 
These cumulants are defined as the ensemble average of $\delta_R^n$ while keeping the long mode
at that Lagrangian position (e.g., $\chiil(\vq)$ in our case) fixed:
%----------
\be
\kappa^{(n)}_R(\vq) \equiv \langle\delta_R^n\rangle_{{\rm c}, s|\ell}.
\ee
%----------
Then, we can easily show how the long mode, which couples with the short mode due to the primordial non-Gaussianities, 
modify the cumulants of the short mode locally.

In Eq.~(\ref{eq:deltas}), we can see that the nonlinear coupling constant in the short mode of the density contrast
can be obtained by replacements of the coupling constants to the effective ones 
%----------
\be
c_{i,j} \to c_{i,j}(\vq) \equiv c_{i,j}+\Delta c_{i,j}(\vq),\label{eq:cij_effective}
\ee
%----------
which vary from position to position proportional to a certain combination of $\chiol$ and $\chitl$.
We can obtain the conditional cumulants by applying the above replacements to Eq.~(\ref{eq:cumulants})
derived for the unconditional cumulants.
Let us consider first the conditional variance of the short mode fluctuations of the density contrast.
We define the local rms fluctuation, $\sigma_R(\vq) \equiv [\kappa^{(2)}_R(\vq)]^{1/2}$, and this is given by
%----------
\be
%\sigma_R(\vx) &=& \left[1+(1-\alpha)\Delta c_{1,0}(\vx)+\alpha\Delta c_{0,1}(\vx)\right]\bar\sigma_R,\nonumber\\
%\displaystyle &=&\left\{1+\left[2(1-\alpha)c_{2,0}+\alpha c_{1,1}\right]\chiol(\vx)+\left[(1-\alpha)c_{1,1}+2\alpha c_{0,2}\right]\chitl(\vx)\right\}\bar\sigma_R,
\sigma_R(\vq) = \Bigl\{1+\bigl[2(1-\alpha)c_{2,0}+\alpha c_{1,1}\bigr]\chiol(\vq)+\bigl[(1-\alpha)c_{1,1}+2\alpha c_{0,2}
\bigr]\chitl(\vq)\Bigr\}\bar\sigma_R\label{eq:conditionalsigmaR},
\ee
%----------
where we keep only linear corrections from $\chiol$ and $\chitl$. 
Similarly, we can compute the higher-order conditional cumulants of the short mode when the long modes are fixed. 
Again, we keep the leading-order corrections arising from $\chiil$, and obtain
%----------
\be
\kappa_R^{(3)}(\vq) &=& \bigl[\fnl+\Delta\fnl(\vq)\bigr]\hat{\kappa}_{R,\fnl}^{(3)},\label{eq:kappa3local}\\
\kappa_R^{(4)}(\vq) &=& 
\bigl[\gnl+\Delta\gnl(\vq)\bigr]\hat{\kappa}_{R,\gnl}^{(4)}+
\bigl[\tnl+\Delta\tnl(\vq)\bigr]\hat{\kappa}_{R,\tnl}^{(4)},
\label{eq:kappa4local}
\ee
%----------
for the skewness and kurtosis. Here we define the effective shifts in the coupling constants, $\Delta\fnl(\vq)$, 
$\Delta\gnl(\vq)$ and $\Delta\tnl(\vq)$. They can be explicitly written down as linear combinations of 
$\chiol(\vq)$ and $\chitl(\vq)$:
%---------------
\be
\Delta\fnl(\vq) &=& \frac53\Bigl\{\Bigl[(1-\alpha)^2(4c_{2,0}^2+3c_{3,0})+\alpha(1-\alpha)(2c_{1,1}c_{2,0}+c_{1,1}^2+2c_{2,1})
\nonumber\\
&&+\alpha^2(2c_{0,2}c_{1,1}+c_{1,2})\Bigr]\chiol(\vq) + \Bigl[1\leftrightarrow2\Bigr]\chitl(\vq)\Bigr\},
\label{eq:deltafnl}\\
\Delta\gnl(\vq) &=& \frac{25}9\Bigl\{\Bigl[2(1-\alpha)^3(3c_{2,0}c_{3,0}+2c_{4,0}) 
+ \alpha(1-\alpha)^2(4c_{2,0}c_{2,1}+c_{1,1}c_{2,1}+3c_{3,1})\nonumber\\
&& + 2\alpha^2(1-\alpha)(c_{2,0}c_{1,2}+c_{1,1}c_{1,2}+c_{2,2}) + \alpha^3(3c_{1,1}c_{0,3}+c_{1,3})\Bigr]\chiol(\vq)\nonumber\\
&& + \Bigl[1\leftrightarrow2\Bigr]\chitl(\vq)\Bigr\},\label{eq:deltagnl}\\
\Delta\tnl(\vq) &=& \Bigl\{\Bigl[8(1-\alpha)^3(2c_{2,0}^3+3c_{2,0}c_{3,0})\nonumber\\
&&+4\alpha(1-\alpha)^2(2c_{2,0}c_{1,1}^2+2c_{2,0}^2c_{1,1}+c_{1,1}c_{2,1}+3c_{1,1}c_{3,0}+2c_{2,0}c_{2,1})\nonumber\\
&&+2\alpha^2(1-\alpha)(c_{1,1}^3+4c_{2,0}c_{1,1}c_{0,2}+2c_{1,1}^2c_{0,2}+2c_{1,1}c_{2,1}+4c_{0,2}c_{2,1}+2c_{1,1}c_{1,2})\nonumber\\
&&+8\alpha^3(c_{1,1}c_{0,2}^2+c_{0,2}c_{1,2})\Bigr]\chiol(\vq) + \Bigl[1\leftrightarrow2\Bigr]\chitl(\vq)\Bigr\},
\label{eq:deltatnl}
\ee
%---------------
where the coefficients for $\chitl(\vx)$, denoted as $[1\leftrightarrow2]$, can be obtained by performing 
permutations $c_{m,n}\leftrightarrow c_{n,m}$ and $\alpha \leftrightarrow (1-\alpha)$ to the corresponding 
coefficients of $\chiol(\vx)$.
See also Appendix~\ref{app:formula} for the results for single-field non-Gaussianities (\ref{eq:Bardeen}).

An important point here is that the $n$th-order correction in Eq.~(\ref{eq:curvature}) modifies the $n$th-order
conditional cumulant, while the unconditional cumulant at the same order is not affected.
For instance, if one is interested in quadratic models with $\fnl\neq0$, 
the first non-trivial contribution to the conditional cumulants appears
in the variance, while the unconditional variance does not depend on $\fnl$. 
Similarly, when $\gnl\neq0$, the conditional skewness has a linear dependence on the parameter $\gnl$
(see also \cite{Smith11b,Desjacques11a}), while this parameter gives the amplitude of the unconditional kurtosis.

%===========================================================
\subsection{Halo number density and clustering}
\label{subsec:halo}
%===========================================================

The halo mass function is known to be almost universal and solely determined 
by the parameter $\nu\equiv \deltac/\sigma_R$, where $\deltac$ denotes the critical density 
above which halos form at $z=0$ \cite{Jenkins01}.
The mass function is usually characterized by $\bar{f}(M,z)$:
%----------
\be
\bar{n}(M,z) &\equiv& \frac{\rmd \mathcal{N}}{\rmd M} = \bar{f}(M,z)\,\frac{\bar{\rho}}{M^2}
\left|\frac{\rmd\ln\bar\sigma_R^{-1}}{\rmd \ln M}\right|, \label{eq:massfunc}
\ee
%----------
where we denote by $\bar{n}(M,z)$ the number density of halos with mass $M$ at redshift $z$, and 
$\bar\rho$ is the cosmic mean density \cite{Press74}.
When the halo mass function is universal, the function $\bar{f}(M,z)$ depends only on the peak height $\nu$.
However, as we are interested in corrections from the higher-order cumulants, we slightly relax the assumption of 
the universality. We allow a dependence on the density cumulants at any order and write
%----------
\be
\bar{f}(M,z) = \bar{f}\left(\bar{\sigma}_R,\bar{\kappa}^{(3)}_R,\bar{\kappa}^{(4)}_R,\ldots;\deltac\right).
\ee
%----------
Here, the argument $R$ has a one-to-one correspondence to the halo mass $M$ via Eq.~(\ref{eq:mass2rad}).

We then assume that the halo number density field is a local process in Lagrangian space: 
given the conditional cumulants defined in the previous subsection, we can deterministically predict the
number of halos at that (Lagrangian) position.
More specifically, we assume that the halo number density is a function of the conditional cumulants smoothed with the
physical scale corresponding to the halo mass in the Lagrangian space. We introduce $f(\vq | M, z)$ 
analogously to its {\it unconditional} counterpart, $\bar{f}$, in Eq.~(\ref{eq:massfunc}), and
assume that this function can be obtained by replacements of the arguments in $\bar{f}$ as 
%----------
\be
\bar\kappa^{(n)}_R \to \kappa^{(n)}_R(\vq), 
\ee
%----------
in addition to the change in the critical density for the formation of halos in Eq.~(\ref{eq:deltac_effective}).
We have
%----------
\be
f(\vq | M,z) = \bar{f}\left(\sigma_R(\vq), \kappa^{(3)}_R(\vq), \kappa^{(4)}_R(\vq), \ldots; \deltac-\delta_\ell(\vq)\right).
\ee
%----------

This function gives us a simple route to the halo power spectrum.
The halo density contrast in the Lagrangian space can be calculated as
%----------
\be
\displaystyle\delta_h^{\rm L}(\vq | M, z) &=& \frac{f(\vq | M, z)}
{\bar{f}(M, z)}-1.
\label{eq:deltahL}
\ee
%----------
By expanding this equation into Taylor series of $\delta_\ell$, $\chiol$ and $\chitl$, and collecting all the 
leading-order contributions, 
we obtain the linear halo overdensity field in the Lagrangian space. We then map this field to the Eulerian space.
At leading order, we have
%----------
\be
\delta_h(\vx) &=& b_\delta\,\delta(\vx) + b_{\chio}\,\chio(\vx) + b_{\chit}\,\chit(\vx),\label{eq:deltah}
\ee
%----------
with coefficients defined through the derivatives of the mass function:
%----------
\be
b_\delta &=& 1-\frac{\partial\ln f}{\partial \deltac},\label{eq:bdelta}\\
b_\chii &=& \frac{\partial\ln f}{\partial\ln\sigma_R}
\frac{\partial\ln\sigma_R}{\partial\chiil} + \sum_{n=3}^\infty\frac{\partial\ln f}{\partial\kappa_R^{(n)}}
\frac{\partial\kappa_R^{(n)}}{\partial\chiil}, \qquad i=1, 2. \label{eq:bchii}
\ee
%----------
In Eq.~(\ref{eq:deltah}), we dropped the subscript $\ell$ for notational convenience.
Note that the first term in Eq.~(\ref{eq:bdelta}) comes from the mapping to the Eulerian space.
Note also that the derivatives of the conditional cumulants with respect to $\chiil$ can be written down explicitly in terms of 
$c_{i,j}$ and $\alpha$ by using the expressions derived in Sec.~\ref{subsubsec:localcumulants}.
Taking the average of the square of Eq.~(\ref{eq:deltah}) in Fourier space,
we finally arrive at the halo power spectrum:
%----------
\be
P_{\rm h}(k) &=& b_\delta^2\,P_{\delta}(k) + 2\,r_{\rm MF}\,b_\delta\,b_{\zeta}\,P_{\delta\zeta}(k) + b_{\zeta}^2\,P_{\zeta}(k),
\label{eq:ph}
\ee
%----------
where the density and density-curvature power spectra are defined as
%----------
\be
P_{\delta}(k) \equiv \mathcal{M}^2(k)P_\zeta(k),\qquad
P_{\delta\zeta}(k) \equiv \Mk P_\zeta(k),
\ee
%----------
and we parameterize the bias coefficients by
%----------
\be
b_\zeta &\equiv& \sqrt{(1-\alpha)b_\chio^2+\alpha b_\chit^2},\\
r_{\rm MF} &\equiv& \frac{(1-\alpha)b_\chio+\alpha b_\chit}{\sqrt{(1-\alpha)b_\chio^2+\alpha b_\chit^2}}.\label{eq:rmf}
\ee
%----------
As we will discuss later in more detail, the parameter $b_\zeta$ characterizes the amplitude of the non-Gaussian
corrections to the halo power spectrum. On the other hand, the other parameter $r_{\rm MF}$ captures the 
shape of the scale dependent bias, and plays a key role to distinguish the models for the primordial non-Gaussianities
based on multi fields from those based on a single field.

It is interesting to see that a wide class of models described by Eq.~(\ref{eq:curvature}) end up 
with the same result for the halo power spectrum with only two parameters responsible for 
non-Gaussian corrections ($b_\zeta$ and $r_{\rm MF}$) in addition to the usual (Gaussian) bias factor, 
$b_\delta$. 
Moreover, our result can be generalized to models with
more than two independent fields very easily.
If we employ $N$ statistically independent Gaussian fields, $\chi_n$, and expand the curvature perturbations as
%----------
\be
\zx = \sum_n\chi_n(\vx)+\sum_{i_1,\cdots,i_N}c_{i_1, \cdots, i_N}\left[\prod_n\chi_n^{i_n}(\vx) - 
\left\langle\prod_n\chi_n^{i_n}\right\rangle\right],
\label{eq:curvature2}
\ee
%----------
the halo density contrast of Eq.~(\ref{eq:deltah}) becomes
%----------
\be
\delta_h(\vx) &=& b_\delta\,\delta(\vx) + \sum_{n}b_{\chi_n}\,\chi_n(\vx),\label{eq:deltah3}
\ee
%----------
with the index $n$ running from unity to $N$.
From this equation, it is clear that Eq.~(\ref{eq:ph}) still holds as long as all the $N$ fields have the same spectral index.
In this case, the bias coefficients are given by
%----------
\be
b_\zeta &\equiv& \sqrt{\sum_n \alpha_n b_{\chi_n}^2},\\
r_{\rm MF} &\equiv& \frac{\sum_n \alpha_n b_{\chi_n}}{\sqrt{\sum_n \alpha_n b_{\chi_n}^2}},\label{eq:rmf_3}
\ee
%----------
with the fraction in the power spectrum amplitude of the $n$-th field denoted as $\alpha_n$ that satisfy
%----------
\be
\sum_n \alpha_n = 1.
\ee
%----------
Equation~(\ref{eq:deltah3}) can be regarded as a generalization of the multi-variate biasing
in case of the single-field models discussed in \cite{Giannantonio10}.

%----------
\begin{figure}[!t]
\begin{center}
\includegraphics[height=7.2cm]{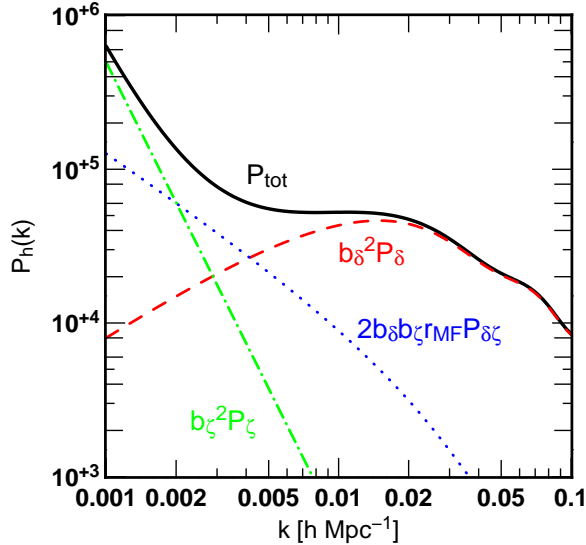}
\caption{Example plot of the halo power spectrum (Eq.~\ref{eq:ph}) in case of $b_\delta = 2$, 
$b_\zeta = 100$ and $r_{\rm MF}=1$. We plot the total power spectrum by solid line,
while the three terms in that equation are respectively shown in dashed, dotted and dot-dashed lines.}
\label{fig:demo0}
\end{center}
\end{figure}
%----------

We show an example halo power spectrum in Fig.~\ref{fig:demo0}.
We can see how each of the three terms in Eq.~(\ref{eq:ph}) contributes to the total power spectrum.
The non-Gaussian corrections are prominent on large scales ($\simlt 0.01 h$Mpc$^{-1}$).
The parameters, $b_\zeta$ and $r_{\rm MF}$ as well as $b_\delta$ can be determined from observed
power spectrum with a sufficient survey volume thanks to the different $k$-dependence of the three terms.
We propose to use these parameters as direct observables from the measurements of the scale-dependent 
halo bias instead of popular non-Gaussian parameters such as $\fnl$ and $\gnl$, because the latter set of
parameters have strong degeneracy in many cases.
We will see how our new parameterization helps us to test different class of models for the primordial 
non-Gaussianities in the next section.

%===========================================================
\subsection{Recovery of previous studies}
\label{subsec:comp}
%===========================================================

It is worth comparing our results with the predictions in the literature.
In the presence of the primordial non-Gaussianities based on a single field, 
it is straightforward to show that $r_{\rm MF} = 1$ or $-1$ 
(see Sec.~\ref{subsec:Cauchy-Schwartz} for more discussion on $r_{\rm MF}$).
Then, Eq.~(\ref{eq:ph}) can be rewritten as
%------
\be
P_{\rm h}(k) &=& [b_\delta + \Delta b(k)]^2 P_\delta(k),\\
\Delta b(k) &\equiv& {\rm sign}(r_{\rm MF})\,b_\zeta\,\mathcal{M}^{-1}(k),
\label{eq:ph2}
\ee
%------
where we denote the sign of $A$ as sign$(A)$.
In this case, one can obtain the scale-dependent bias in the
halo power spectrum simply by replacing a constant bias factor with a $k$-dependent one:
$b_\delta \to b_\delta + \Delta b(k)$.
All the effects from the primordial non-Gaussianities are encoded in $b_\zeta$, the amplitude of the
scale-dependent bias.
This parameter in case of the single-field primordial non-Gaussianities (\ref{eq:Bardeen}) 
is explicitly given in Appendix~\ref{app:formula}.

Let us further restrict the model to the quadratic non-Gaussianities (i.e., $\fnl$-type).
We find that the first term in Eq.~(\ref{eq:bchii}) is the dominant contribution to $b_\zeta$, and 
the derivatives of cumulants $\kappa_R^{(n)}$ with $n\ge3$ are small. 
By dropping these small corrections, 
and using that the derivative with respect to $\deltac$ in Eq.~(\ref{eq:bdelta}) can be converted to 
that with respect to $\sigma_R$ under the assumption of the universal halo mass function, we recover 
%------
\be
\Delta b(k) = \frac{6}5\,\fnl\,\deltac\,(b_\delta - 1)\,\mathcal{M}^{-1}(k),
\label{eq:Deltabk}
\ee
%------
derived by \cite{Dalal08}. 
The correction terms from higher-order cumulants might be important if one needs a very accurate 
modeling for the scale-dependent bias.
However, since these terms strongly depends on the non-Gaussian mass function and 
we adopted a rather simplified one described in Appendix~\ref{app:massfunction} in this paper, 
we left further discussions on the accurate modeling in paper II.

We next consider cubic non-Gaussianities described by $\gnl$. 
In this case, the first term of Eq.~(\ref{eq:bchii}) is zero,
and thus the dominant effect comes from the derivative of the mass function with respect to the skewness.
Using the expression for the conditional skewness given in Appendix~\ref{app:formula}, and dropping
the corrections from higher-order cumulants, we have
%------
\be
\Delta b(k) = \frac{9}5\,\gnl\,\hat\kappa^{(3)}_{\fnl,R}\,\frac{\partial\ln f}{\partial\ln \kappa^{(3)}_R}\,\mathcal{M}^{-1}(k),
\label{eq:Deltabk_gnl}
\ee
%------
This is identical to Ref.~\cite{Smith11b}.
One can recover the formula for the scale-dependent bias derived in \cite{Desjacques11a,Desjacques11b} by 
substituting the mass function based on the Edgeworth expansion to this expression (see Appendix~\ref{app:massfunction}). 
Indeed, Ref.~\cite{Smith11b} demonstrated that the $N$-body data can be 
accurately explained by evaluating the derivative of the mass function directly from simulations.

Let us put some comments on non-Gaussianities based on multi-field initial conditions.
Refs.~\cite{Tseliakhovich10,Smith11a} compute the clustering of halos in a two-field inflationary model 
using the peak-background split.
They employ two fields, one is Gaussian and the other is non-Gaussian with a quadratic correction (i.e., $\fnl$-type),
and thus our model (\ref{eq:curvature}) includes their model.
However, their main focus is on the stochasticity between the halo and the underlying matter density fields.
Although our derivation in this paper is equivalent to theirs, our focus is more on the auto-power spectrum of halos.
It is straightforward to show that the stochasticity arises between the halo and the matter fields 
in our models when $\alpha\neq0$ regardless of the order of the non-Gaussian corrections, and this is another
unique feature to test multi-field non-Gaussianities.

Note also that our analytical calculation of the halo power spectrum
is valid only at the linear order. Although we are interested in the clustering on large scales 
(say, $\simgt1h^{-1}$Gpc) and linear theory is expected to work well, any nonlinearity (e.g., nonlinear gravitational 
growth or nonlinear bias) can in principle modify our results.
For example, the nonlinearity in the halo bias can source a similar scale dependence as shown in \cite{Matarrese08}
(and see also \cite{Desjacques11a} for a comparison among different derivations of the scale-dependent bias
including the effect of nonlinear bias in the thresholding scheme).
As another example, the matter power spectrum is affected by a loop correction in case of $\gnl$-type models, which
does not vanish at large scale limit \cite{Desjacques10a}.
Another important assumption in our analysis is the locality of the formation of halos in Lagrangian space.
These issues will be discussed in paper II with fully nonlinear treatment based on cosmological 
$N$-body simulations.

%===========================================================
\section{Testing non-Gaussian models with observation}
\label{sec:obs}
%===========================================================
We have shown that 
%the local-type non-Gaussian models described as a Taylor expansion
%of the curvature perturbations, $\zeta$, into an arbitrary number of Gaussian fields
our model (\ref{eq:curvature2}) generally results in the halo power spectrum Eq.~(\ref{eq:ph}).
The final result of the halo power spectrum contains only three parameters, $b_\delta$, $b_\zeta$
and $r_{\rm MF}$, and they can be determined or constrained directly from the observation using
the different $k$-dependence of the three terms in Eq.~(\ref{eq:ph}).
Following this result, we discuss what properties of the primordial non-Gaussianities can be tested 
by measurements of these parameters.

%===========================================================
\subsection{Cauchy-Schwartz inequality and multi-field non-Gaussianities}
\label{subsec:Cauchy-Schwartz}
%===========================================================
We discuss what we can learn from the parameter $r_{\rm MF}$ defined in Eq.~(\ref{eq:rmf_3}).
By simple calculation, one can show that $r_{\rm MF}$ satisfies the following inequality:
%----------
\be
\left|r_{\rm MF}\right| \leq 1,\label{eq:inequality}
\ee
%----------
where the equality holds when the curvature perturbations originate from a single field.
In multi-field models, the above equality holds only when all the coefficients $b_\chii$ are the same.
Note that this inequality is the Cauchy-Schwartz inequality among Gaussian fields, $\chii$.

Measurement of this parameter opens a new window to distinguish between single-field and multi-field 
primordial non-Gaussianities.
If the absolute value of $r_{\rm MF}$ measured from the observations of the galaxy clustering 
significantly deviates from unity,
we can rule out single-field models of the primordial fluctuations.
Interestingly, this inequality is shown to be identical to the Suyama-Yamaguchi inequality in case
of the quadratic non-Gaussianities under some approximations.
Let us focus on two-field models (\ref{eq:curvature}) for simplicity.
As we have already mentioned, the dominant contribution to the bias coefficients in Eq.~(\ref{eq:bchii})
is the first term in the presence of quadratic non-Gaussianities, $c_{i,j}$ with $i+j=2$.
If we drop the higher-order corrections, we have
%----------
\be
b_\chio &\approx& \left[2(1-\alpha)c_{2,0}+\alpha c_{1,1}\right]\frac{\partial\ln f}{\partial\ln\sigma_R},\label{eq:bchi1_approx}\\
b_\chit &\approx& \left[(1-\alpha)c_{1,1}+2\alpha c_{0,2}\right]\frac{\partial\ln f}{\partial\ln\sigma_R}.\label{eq:bchi2_approx}
\ee
%----------
Substituting these expressions into Eq.~(\ref{eq:rmf}), we finally have
%----------
\be
r_{\rm MF} \approx \frac{6\fnl}{5\sqrt{\tnl}},\label{eq:rmf_approx}
\ee
%----------
with a help of the definitions of $\fnl$ and $\tnl$ in Eqs.~(\ref{eq:fnl}) and (\ref{eq:tnl}).
Note that Eq.~(\ref{eq:rmf_approx}) still holds in more general case of Eq.~(\ref{eq:curvature2}).
Thus, we can directly test the Suyama-Yamaguchi inequality from the observed power spectrum.
Note that our parameter $r_{\rm MF}$ is equivalent to $A_{\rm NL}^{-1/2}$ introduced in
\cite{Smidt10} under this approximation.
Since our inequality is valid for models with higher-order non-Gaussianities, it can be regarded as a generalization
of the Suyama-Yamaguchi inequality, and can be directly tested from the observed halo power spectrum.

%----------
\begin{figure}[!t]
\begin{center}
\includegraphics[height=7.2cm]{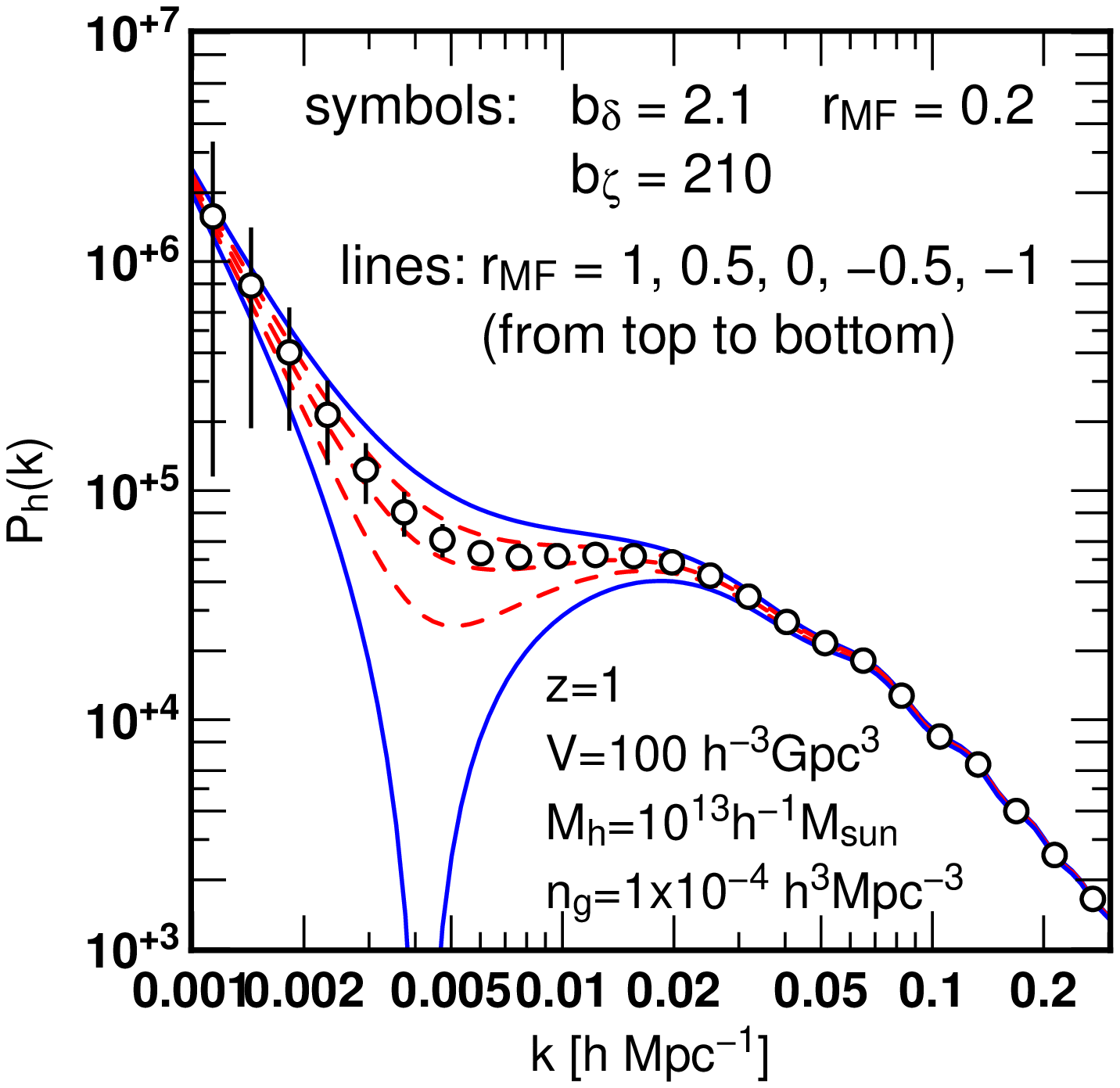}
\includegraphics[height=7.2cm]{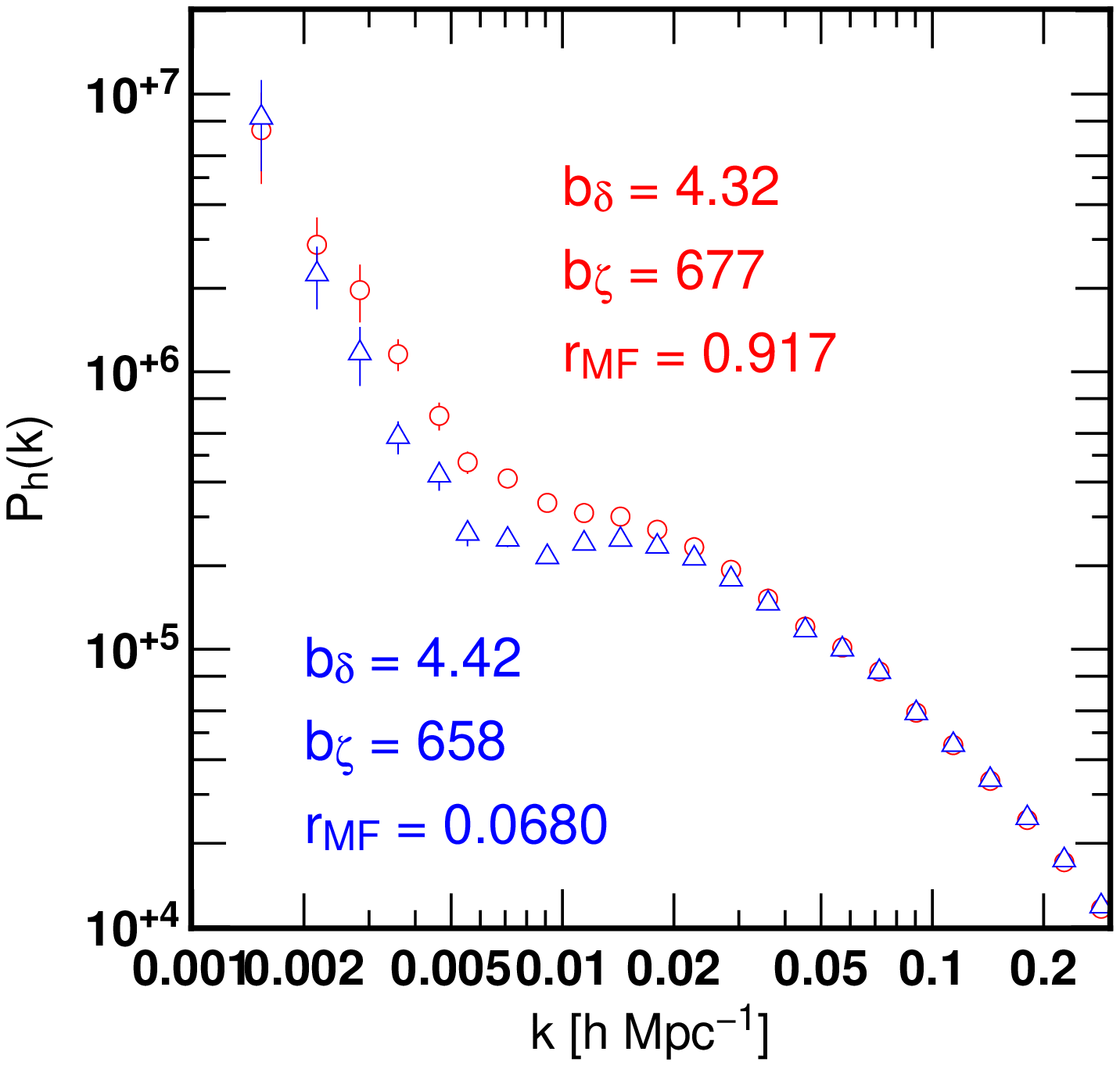}
\caption{Dependence on the shape parameter $r_{\rm MF}$ of the halo power spectrum.
Left: analytical results of the $r_{\rm MF}$ dependence at $z=1$. Lines show models with 
different $r_{\rm MF}$ while $b_\delta=2.1$ and $b_\zeta=210$ are fixed.
Expected error bars on the galaxy power spectrum from a $100\,h^{-3}$Gpc$^3$ class survey are also shown
assuming $r_{\rm MF}=0.2$ as the fiducial value. 
We assume that the typical host halo mass is $10^{13}\,h^{-1}M_\odot$ 
(this gives the value of $b_\delta$ shown in the figure), and
the number density of halos is $n_{\rm h} = 10^{-4}\,h^3$Mpc$^{-3}$.
Right: examples for halo power spectra measured from $N$-body simulations. 
We fit the data by model (\ref{eq:ph}) to find the best-fit parameters, $b_\delta$, $b_\zeta$ and $r_{\rm MF}$.
They are shown in the figure for each model.}
\label{fig:demo1}
\end{center}
\end{figure}
%----------

The parameter $r_{\rm MF}$ controls the shape of the scale dependent bias.
This is illustrated in Fig.~\ref{fig:demo1}. 
In the left panel, we show how the halo power spectrum depends on $r_{\rm MF}$ when $b_\delta$ and $b_\zeta$ are fixed.
The solid lines correspond to models of primordial non-Gaussianities originating from a single-field with $\fnl=\pm 100$, 
while the dashed lines show the halo power spectra from multi-field non-Gaussian initial conditions.
We also plot by symbols the expected error bars from a $100h^{-3}$Gpc$^3$ survey at $z=1$.
We employ the formula by \cite{Feldman94} for the $1$-$\sigma$ error bars that is valid for Gaussian fields:
%----------
\be
[\Delta P_g(k)]^2 = \frac{1}{N_{\rm mode}}\left[P_g(k) + \frac{1}{n_g}\right]^2,
\ee
%----------
where $N_{\rm mode}$ denotes the number of independent Fourier modes.
The fiducial model shown as the error bars in the left panel corresponds to 
$\fnl = 20$ and $\tnl = 14,400$, if we assume that the non-Gaussianities are quadratic.
The value of $\fnl$ is small enough to satisfy the current observational bounds, while that of $\tnl$ 
is much larger than $(36/25)\fnl^2$, which might be possible to detect from future observations.

From the figure, we can see the largest difference among the five lines at $k\sim 0.005h$Mpc$^{-1}$,
where the $P_{\delta\zeta}$ term in Eq.~(\ref{eq:ph}) has the largest impact.
On the other hand, the they converge at the small-$k$ limit since we adopt the same value 
of $b_\zeta$ for them. 
We need to measure the broadband shape of the power spectrum covering this entire scale at $\simgt h^{-1}$Gpc
to obtain meaningful constraints on $r_{\rm MF}$.
The planned surveys such as EUCLID \cite{EUCLID} can measure the galaxy power spectrum quite accurately 
even at such a large scale thanks to their huge survey volumes.
They will provide great opportunities for this kind of tests.

We also show in the right panel of Fig.~\ref{fig:demo1} the measured halo power spectra from 
$N$-body simulations at $z=1$. We plot the results obtained in simulations starting with two different 
non-Gaussian initial conditions; a single-field model (circles), and a two-field model (triangles).
The input parameters of these simulations are $c_{2,0}=60$ and $\alpha=0$ for the single-field model, and
$c_{2,0}=120$, $c_{0,2}=-120$ and $\alpha=1/2$ for the two-field model.
The derived non-Gaussian parameters for the two simulations are $(\fnl,\,\tnl) = (100,\,14400)$ and $(0,\,14400)$,
respectively.
Thus, the two simulations are expected to have roughly the same values for $b_\delta$ and $b_\zeta$.
We fit the measured halo power spectrum by Eq.~(\ref{eq:ph}) to obtain $b_\delta$, $b_\zeta$ and $r_{\rm MF}$,
which are shown in the panel.
The best-fit value of $r_{\rm MF}$ in case of the single-field model shown in the figure is consistent with unity
within the statistical uncertainty, as expected.
On the other hand, the multi-field model clearly disfavors $|r_{\rm MF}| = 1$.
These results basically confirm our analytical expectations.
See paper II for more detail on the simulations and analyses.

It is interesting to note that in Ref.~\cite{Gong11}, the authors reach a similar conclusion for testing $\fnl$ and $\tnl$ through
the shape of the halo power spectrum but from a different derivation based on the clustering of peaks 
above a threshold. Our results agree with theirs qualitatively. One advantage of our formulation is 
that our inequality~(\ref{eq:inequality}) is always valid with non-Gaussian corrections at an arbitrary order 
and can be used to test multi-field models for the primordial fluctuations in general.

%===========================================================
\subsection{Mass dependence and the order of the non-Gaussian corrections}
\label{subsec:mass}
%===========================================================
We finally discuss how we should distinguish the quadratic non-Gaussianities (i.e., $\fnl$-type) from
higher-order models.
As we have shown in the previous subsection, the shape of the halo (or galaxy) power spectrum on large scales
only gives the information about the number of independent fields through the parameter $r_{\rm MF}$.
However, the $\fnl$-type non-Gaussianities are distinguishable from higher-order ones
if one examines more than one galaxy populations with different bias parameters, in principle.
In this section, we discuss how the ratio
%-----------------
\be
A_{\rm nG}(M) \equiv \frac{b_\zeta(M)}{b_\delta(M) - 1},
\ee
%-----------------
can be used for this test.

\begin{figure}[!t]
\begin{center}
\includegraphics[height=7.29cm]{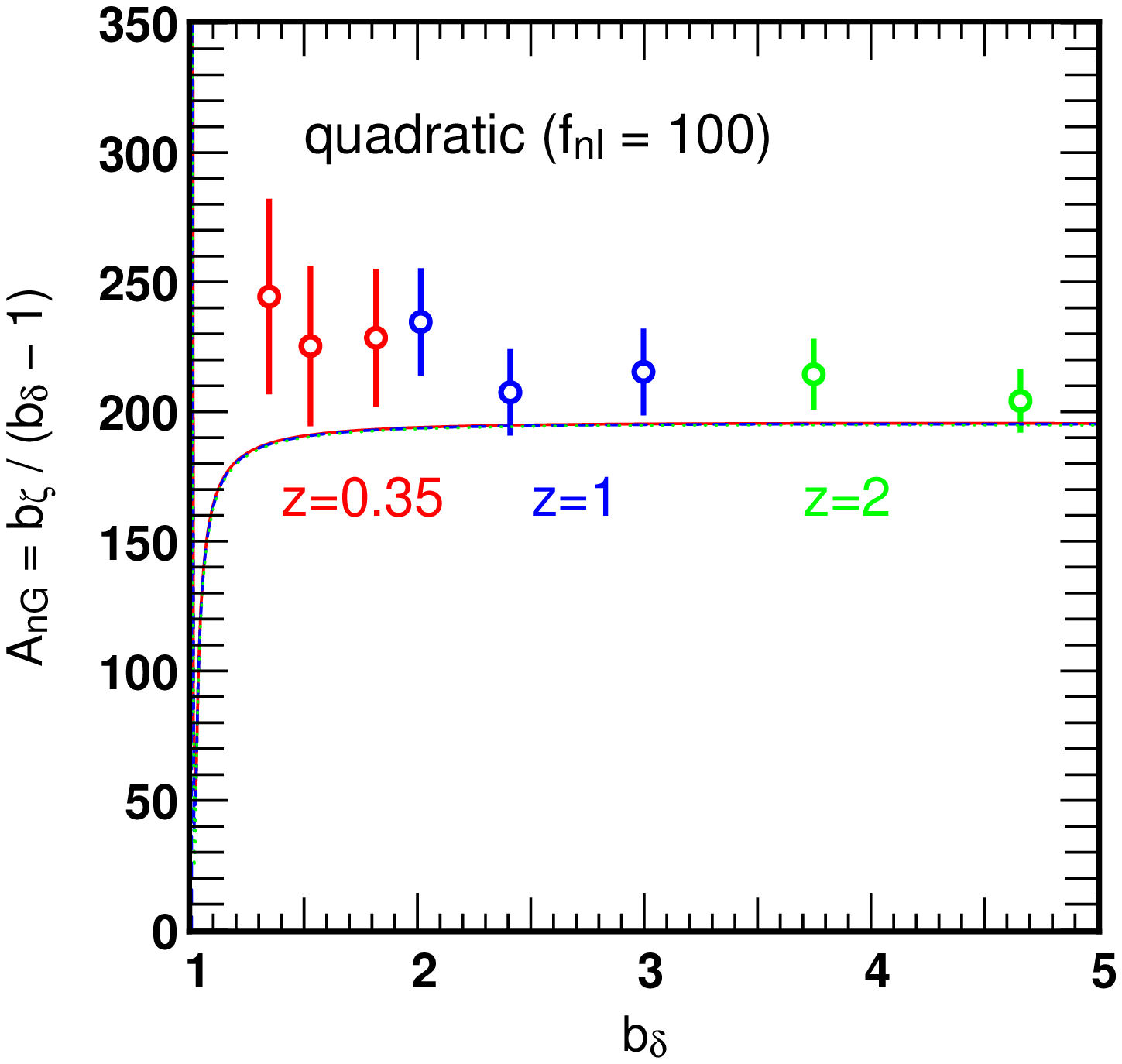}
\includegraphics[height=7.2cm]{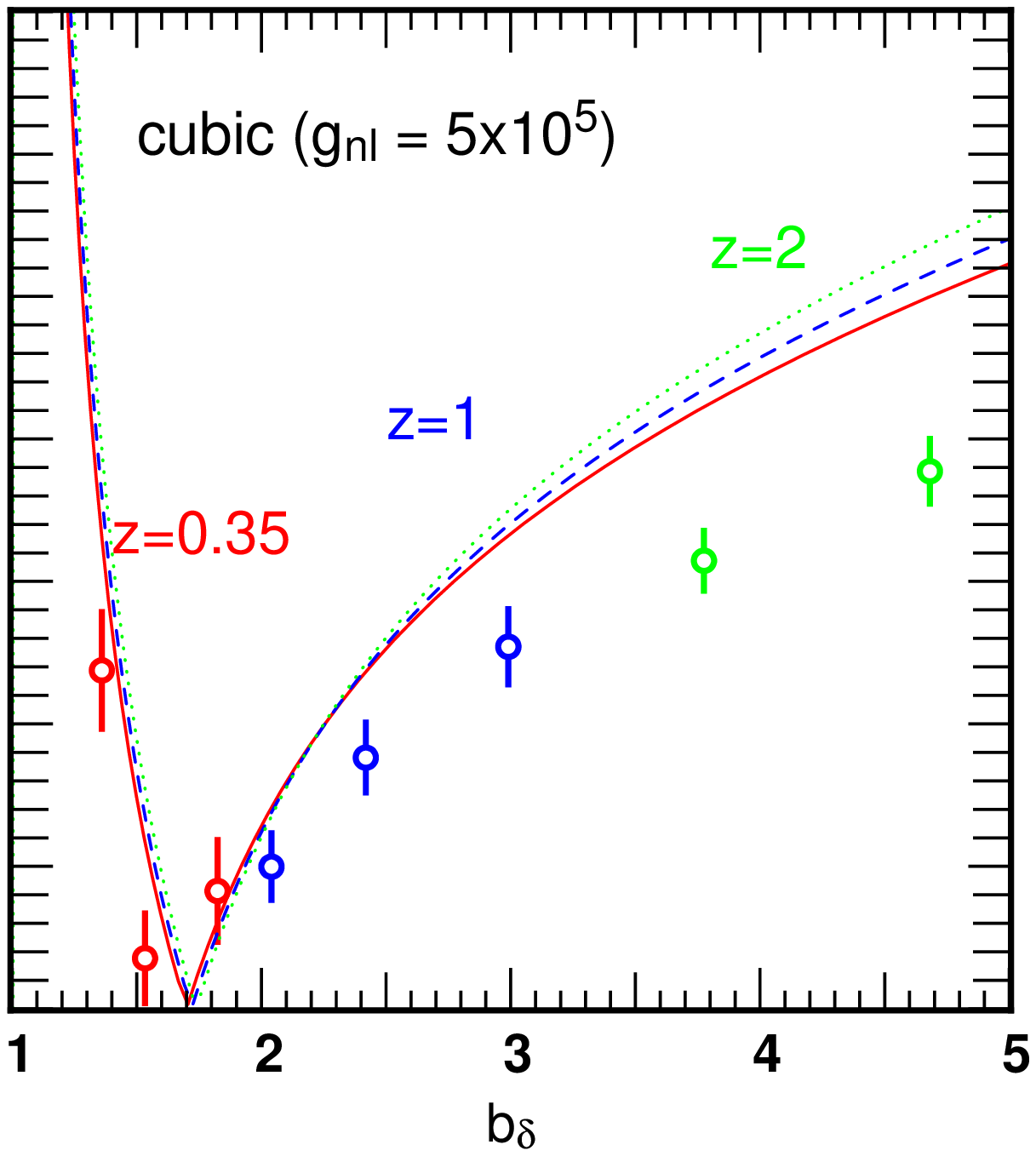}
\caption{Dependence on the Gaussian bias factor, $b_\delta$, of the non-Gaussian correction to the halo bias.
We plot the combination $A_{\rm nG} = b_\zeta / (b_\delta - 1)$ as a function of $b_\delta$ at redshifts $0.35$,
$1$ and $2$. 
The error bars show the $1$-$\sigma$ uncertainty obtained by fitting to $N$-body simulations,
while the lines depict our theoretical predictions at $z=0.35$ (solid), $z=1$ (dashed) and $z=2$ (dotted).
The left (right) panel plots the results for quadratic (cubic) non-Gaussianities characterized by $\fnl$ ($\gnl$).
}
\label{fig:demo2}
\end{center}
\end{figure}

For simplicity, let us focus on the primordial curvature perturbations originating from a single Gaussian 
field (\ref{eq:Bardeen}). 
Figure~\ref{fig:demo2} shows how the amplitude of the 
non-Gaussian correction to the halo bias scales with the Gaussian bias factor for halo catalogs with 
different masses at different redshifts. 
We plot $A_{\rm nG}$ measured from $N$-body simulations (error bars) and estimated by our analytic calculation (lines).
We adopt two models for the primordial non-Gaussianities, and show the results for $\fnl=100$ in the left panel 
and $\gnl=5\times10^5$ in the right panel.

%----------
%----------

As expected from Eq.~(\ref{eq:Deltabk}) derived in the literature, this ratio remains almost constant over the Gaussian 
bias factor (or the halo mass) and redshift in case of quadratic non-Gaussianities. 
In contrast, the scaling of $A_{\rm nG}$ with $b_\delta$ observed in the right panel is
very different. A clear increasing tendency is visible from both simulations (symbols) and analytical predictions (lines).
Interestingly, the redshift dependence of the relation between $A_{\rm nG}$ and $b_\delta$ is very weak similarly to the
left panel for the $\fnl$-type primordial non-Gaussianities.
Since we need an accurate modeling of the derivatives of the mass function with respect to the cumulants
to compute the analytical curves, it is generally difficult to reach a quantitative agreement with simulations.
See Appendix~\ref{app:massfunction} for more detail of the mass function we adopt in computing the 
analytical predictions. 

Although our analytical calculation has a slight offset to the $N$-body data, the qualitative agreement 
with the $N$-body simulations is rather encouraging.
The difference between the two models shown in Fig.~\ref{fig:demo2} originates from the different origins 
of the scale-dependent bias.
In the presence of a non-zero $\fnl$, most of the effect comes from the conditional variance as was shown 
in the previous section.
In this case, $b_\zeta$ roughly scales as $\propto (b_\delta-1)$, and thus $A_{\rm nG}$ does not depend on the halo mass.
In contrast, we find that the origin of the scale-dependent bias is in the conditional skewness in the presence 
of non-zero $\gnl$. 

Using this feature, we propose the following test. 
Measure the power spectrum from multiple tracers, such as blue and red galaxies, 
massive and less massive halos at any redshifts.
Then, fit by Eq.~(\ref{eq:ph}) to obtain $b_\zeta$ as well as $b_\delta$ and compare the ratio $A_{\rm nG}$ 
measured from each tracer.
If the two (or more) values of $A_{\rm nG}$ are inconsistent with statistical significance, 
that might be a sign of the primordial non-Gaussianities which originate from higher-order coupling (i.e., $\gnl$ or even higher).
We should have a constant $A_{\rm nG}$ independent of the properties of the tracers in case of the 
quadratic non-Gaussianities characterized by $\fnl$.
We also examine the parameter $A_{\rm nG}$ in the presence of quartic-order non-Gaussianities described by 
$\hnl$ (see Eq.~\ref{eq:Bardeen}). We find that the $b_\delta$ dependence of this parameter is 
similar to what is found in the presence of cubic non-Gaussianities.
Thus, we conclude that the constancy of $A_{\rm nG}$ is generally a key feature for discriminating $\fnl$-type 
non-Gaussianities from higher-order models.

%===========================================================
\section{Summary and Discussions}
\label{sec:summary}
%===========================================================

We computed the halo power spectrum in a class of local-type non-Gaussianities
described by Eq.~(\ref{eq:curvature2}) based on the peak-background split formalism.
We found that the resultant halo power spectrum generally exhibits a scale dependence in the
large scale bias factor and is parameterized by three
parameters; the ordinary Gaussian bias factor, $b_\delta$, the shape parameter of
the scale dependence, $r_{\rm MF}$, and the amplitude of the scale dependence, $b_\zeta$.
We showed how one can distinguish between non-Gaussianities based on a single field and multi fields 
through the measurement of $r_{\rm MF}$. This allows us a direct test of the generalized Suyama-Yamaguchi
inequality (\ref{eq:inequality}) from observational data. 
Another parameter, $b_\zeta$ might be useful to distinguish $\fnl$-type models from $\gnl$-type or 
models with even higher-order non-Gaussianities. The parameter $A_{\rm nG}$, 
a combination of $b_\zeta$ and $b_\delta$, is shown to be
almost constant in case of $\fnl$-type non-Gaussianities, though it can have a strong dependence on the 
mass of halos in general. Thus, we propose to measure this parameter from multiple tracers to search for a 
possible dependence of $A_{\rm nG}$ on the halo mass to test higher-order non-Gaussian models.

There are several concerns on the systematic error and possible degeneracy with other models for
the primordial non-Gaussianities.
Since our analytic model is based on the leading-order calculation, any loop-correction can be a source of
systematic error. This can be a problem for both of our two tests, if the correction is large.
We thus have to test our model against simulations.
A detailed comparison with $N$-body simulations will be presented in a separate paper. 

Another systematic effect, though it is avoidable by a careful theoretical modeling,
is induced from a mass dependence in the critical density for the halo formation, so called the 
moving barrier \cite{Sheth02}.
According to the moving barrier argument, the critical density $\deltac$ might not be a constant, 
and should be replaced by $\deltac(M)$ with an explicit halo-mass dependence.
In the presence of non-zero $\fnl$, the ratio $A_{\rm nG}$ can be approximated as
%----------
\be
A_{\rm nG}(M) \approx \frac65\,\fnl\,\deltac(M),
\ee
%----------
when we neglect terms from higher-order cumulants. This equation implies that 
the mass dependence in $\deltac(M)$ can be misinterpreted as a sign of higher-order non-Gaussianities.
We will discuss in paper II how this matters using $N$-body simulations, and show that
in the presence of non-zero $\fnl$, we have no clear evidence of the scale dependence in $\deltac(M)$.

The halo assembly bias is another source of systematics. Ref.~\cite{Reid10} showed
that the amplitude of the scale-dependent bias for a given halo mass can be different
when they divide the halo sample into subsamples with different halo assembly histories.
This can be a problem when the observed galaxies form in halos with a certain merging history preferentially.
The problem of non-Gaussian halo assembly bias may also be interpreted as a dependence on the assembly history of 
the barrier function $\deltac(M)$. We need a careful modeling of the barrier function as a function of the 
assembly history in addition to the mass.

Finally, scale dependent $\fnl$ models can also result in a mass dependent ratio $A_{\rm nG}$. 
In Ref.~\cite{Shandera11}, the authors discuss the scale-dependent bias in models with $k$ dependence in the coupling
parameter $\fnl(k)$. They show that the magnitude of the non-Gaussian correction to the bias in the power spectra 
of massive and less massive halos depends on $\fnl(k)$ evaluated at wavenumber roughly corresponding to the mass 
scale of the halos (see also \cite{Becker11,Desjacques11a,Desjacques11b}). 
Thus, this dependence can mimic the mass dependence of $A_{\rm nG}$ that we found in the presence of 
higher-order non-Gaussianities.
However, by detecting a mass dependence of this parameter, we might be able to falsify 
the simplest non-Gaussian model with a constant $\fnl$, and thus this test is still interesting.

This work is the first step towards establishing a methodology to test non-standard primordial non-Gaussianities
beyond the so-called $\fnl$ type in the era of $100h^{-3}$Gpc$^3$ class full sky surveys.
We expect that our tests for the multi-field non-Gaussianities as well as the higher-order ones will be
possible with future surveys such as EUCLID.

%-----------------------------------------------------------
\acknowledgments
We appreciate Atsushi Taruya for careful reading of the manuscript and giving us valuable
suggestions and comments. We thank Fabian Schmidt for comments on 
the derivative terms in the non-Gaussian halo mass functions.
We also thank Masahide Yamaguchi for useful comments.
T.~N. is supported by a Grant-in-Aid for Japan Society for the Promotion of Science 
(JSPS) Fellows (PD: 22-181) and by World Premier International Research Center 
Initiative (WPI Initiative), MEXT, Japan. Numerical computations for the present work 
have been carried out in part on Cray XT4 at Center for Computational Astrophysics, 
CfCA, of National Astronomical Observatory of Japan, and in part under the 
Interdisciplinary Computational Science Program in Center for Computational Sciences,
University of Tsukuba.
%-----------------------------------------------------------

\appendix

%===========================================================
\section{Fifth-order statistics}
\label{app:quartic}
%===========================================================
In this appendix, we show the relevant formulae for fifth-order statistics in model (\ref{eq:curvature2}).
In actual calculation of the bias coefficients (\ref{eq:bchii}) in the text, we truncate the series at the fifth-order cumulants.
The numerical results shown in Fig.~\ref{fig:demo2} are based on the formulae in this section.

First, the tetraspectrum of the curvature perturbations reads
%---------------
\be
T^{(5)}_\zeta(\vk_1,\cdots,\vk_5) &=& \frac{648}{125}\,\hnl\left[P_\zeta(k_1)P_\zeta(k_2)P_\zeta(k_3)P_\zeta(k_4)+\perm{4}\right]
\nonumber\\
&& + \tnl^{\rm (II)} \left[P_\zeta(k_1)P_\zeta(k_2)P_\zeta(k_3)P_\zeta(|\vk_1+\vk_4|)+\perm{59}\right]
\nonumber\\
&& + \tnl^{\rm (III)} \left[P_\zeta(k_1)P_\zeta(k_2)P_\zeta(|\vk_1+\vk_3|)P_\zeta(|\vk_2+\vk_4|)+\perm{59}\right].
\ee
%---------------
In the above, the coefficients, $\tnl^{\rm (II)}$ and $\tnl^{\rm (III)}$ are respectively given as
%---------------
\be
\tnl^{\rm (II)} = \frac{324}{125}\fnl\,\gnl,\qquad
\tnl^{\rm (III)} = \frac{216}{125}\fnl^3,
\ee
%---------------
when $\zeta$ originates from a single field as in Eq.~(\ref{eq:Bardeen}), 
and they can be different in multi-field cases.
Corresponding to the three terms for the tetraspectrum, we give the fitting formulae for the fifth-order 
cumulant of the smoothed density field at $z=0$:
%---------------
\be
\bar{\kappa}^{(5)}_{R} &=& \hnl\,\hat{\kappa}^{(5)}_{R, \hnl} + \tnl^{\rm (II)}\,\hat{\kappa}^{(5)}_{R, \tnl^{\rm (II)}}
+ \tnl^{\rm (III)}\,\hat{\kappa}^{(5)}_{R, \tnl^{\rm (III)}},\\
\hat{\kappa}^{(5)}_{R, \hnl} &=&1.26\times10^{-11}\bar{\sigma}_R^{5.456},\label{eq:fit_hnl}\\
\hat{\kappa}^{(5)}_{R,\tnl^{\rm (II)}} &=& 7.82\times10^{-11}\bar{\sigma}_R^{5.385},\label{eq:fit_tnl2}\\
\hat{\kappa}^{(5)}_{R,\tnl^{\rm (III)}} &=& 9.67\times10^{-11}\bar{\sigma}_R^{5.288}.\label{eq:fit_tnl3}
\ee
%---------------
The fifth-order conditional cumulants are summarized in the next appendix.

%===========================================================
\section{Exact expressions for single-field non-Gaussianities}
\label{app:formula}
%===========================================================
We mainly focus on non-Gaussian curvature perturbations based on multi fields in the text.
It might be convenient to show the relevant formulae for the single field model (\ref{eq:Bardeen}).

We start with the conditional cumulants. 
First, substituting $\alpha=0$ and with a help of Eq.~(\ref{eq:fnl}), 
the conditional variance of the short-mode fluctuations in Eq.~(\ref{eq:conditionalsigmaR}) is obtained as:
%----------
\be
\sigma_R(\vq) &=& \left[1+\frac{6}5\fnl\zeta_\ell(\vq)\right]\bar\sigma_R,
\ee
%----------
where we use the fact that $\zeta=\chio$ at linear order.
This is equivalent to the results of earlier works in case of $\fnl$-type non-Gaussianities (e.g., \cite{Slosar08,Giannantonio10}).
We then write down the shift in the effective nonlinear parameters Eqs.~(\ref{eq:deltafnl}), 
(\ref{eq:deltagnl}) and (\ref{eq:deltatnl}).
Substituting $\alpha=0$, they are simplified as
%---------------
\be
\Delta\fnl(\vq) &=& \frac35\Bigl(4\fnl^2+3\gnl\Bigr)\zeta_{{\rm G}, \ell}(\vq),\\
\Delta\gnl(\vq) &=& \frac65\Bigl(3\fnl\gnl+2\hnl\Bigr)\zeta_{{\rm G}, \ell}(\vq),\\
\Delta\tnl(\vq) &=&\frac{216}{125}\fnl\Bigl(2\fnl^2+3\gnl\Bigr)\zeta_{{\rm G}, \ell}(\vq).
\ee
%---------------
We can obtain similar expressions for the fifth-order cumulant, $\kappa^{(5)}_R$.
This is given as
%---------------
\be
\kappa^{(5)}_R(\vq) = \left[\hnl + \Delta\hnl(\vq)\right]\hat{\kappa}^{(5)}_{R,\hnl}
+\left[\tnl^{\rm (II)} + \Delta\tnl^{\rm (II)}(\vq)\right]\hat{\kappa}^{(5)}_{R,\tnl^{\rm (II)}}
+\left[\tnl^{\rm (III)} + \Delta\tnl^{\rm (III)}(\vq)\right]\hat{\kappa}^{(5)}_{R,\tnl^{\rm (III)}},
\ee
%---------------
where
%---------------
\be
\Delta\hnl(\vq) &=& \frac35\Bigl(8\fnl\hnl+5\inl\Bigr)\zeta_{{\rm G}, \ell}(\vq),\\
\Delta\tnl^{\rm(II)}(\vq) &=&\frac{972}{625}\Bigl(4\fnl\hnl+3\gnl^2+6\fnl^2\gnl\Bigr)\zeta_{{\rm G}, \ell}(\vq),\\
\Delta\tnl^{\rm(III)}(\vq) &=& \frac{648}{625}\fnl^2\Bigl(9\gnl+4\fnl^2\Bigr)\zeta_{{\rm G}, \ell}(\vq).
\ee
%---------------

Given the formulae for the conditional cumulants up to the fifth order, it is straightforward to
derive the bias coefficient, $b_\zeta$.
It is 
%---------------
\be
{\rm sign}(r_{\rm MF})b_\zeta &=& \frac{6}{5}\fnl\frac{\partial\ln f}{\partial\ln\sigma_R} 
+ \frac{3}{5}\Bigl(4\fnl^2+3\gnl\Bigr)\hat\kappa^{(3)}_{R,\fnl}
\frac{\partial\ln f}{\partial\kappa^{(3)}_R}\nonumber\\
&&+\left[
\frac{6}{5}\Bigl(3\fnl\gnl+2\hnl\Bigr)\hat\kappa^{(4)}_{R,\gnl}
+\frac{216}{125}\fnl\Bigl(2\fnl^2+3\gnl\Bigr)\hat\kappa^{(4)}_{R,\tnl}
\right]\frac{\partial\ln f}{\partial\kappa^{(4)}_R}\nonumber\\
&&+\left[
\frac{3}{5}\Bigl(8\fnl\hnl+5\inl\Bigr)\hat\kappa^{(5)}_{R,\hnl}
+\frac{972}{625}\Bigl(4\fnl\hnl+3\gnl^2+6\fnl^2\gnl\Bigr)\hat\kappa^{(5)}_{R,\tnl^{\rm (II)}}\right.\nonumber\\
&&\left.+\frac{648}{625}\fnl^2\Bigl(9\gnl+4\fnl^2\Bigr)\hat\kappa^{(5)}_{R,\tnl^{\rm (III)}}
\right]\frac{\partial\ln f}{\partial\kappa^{(5)}_R},
\ee
%---------------
where the mass function is give in the next appendix.

%===========================================================
\section{Non-Gaussian halo mass function}
\label{app:massfunction}
%===========================================================
We need knowledge of the mass function to evaluate the bias coefficients.
In this paper, we assume a simple mass function and that is described in this appendix.

We assume that the primordial non-Gaussianities are small, and expand the non-Gaussian 
probability distribution function around normal distribution to obtain Edgeworth series as in \cite{LoVerde08}:
%----------
%\be
%\frac{{\rm PDF}_{\rm nG}(\nu)}{{\rm PDF}_{\rm G}(\nu)}
% &=& 1 + \frac16\frac{\bar\kappa^{(3)}_R}{\bar\sigma^3_R}H_3(\nu)\nonumber\\
%&&+\frac{1}{24}\frac{\bar\kappa^{(4)}_R}{\bar\sigma_R^4}H_4(\nu)+
%\frac1{72}\frac{(\bar\kappa^{(3)}_R)^2}{\bar\sigma_R^6} H_6(\nu)
%\nonumber\\
%&&+ \frac{1}{120}\frac{\bar\kappa^{(5)}_R}{\bar\sigma_R^5}H_5(\nu) 
%+ \frac1{144}\frac{\bar\kappa^{(3)}_R\bar\kappa^{(4)}_R}{\bar\sigma_R^7} H_7(\nu) + \frac1{1296}\frac{(\bar\kappa^{(3)}_R)^3}{\bar\sigma_R^9}H_9(\nu)
%+\ldots,\label{eq:Edgeworth}
%\ee
%----------
%----------
\be
\frac{{\rm PDF}_{\rm nG}(\nu)}{{\rm PDF}_{\rm G}(\nu)}
 &=& 1 + \frac16\overline{C}^{(3)}_RH_3(\nu)\nonumber\\
&&+\frac{1}{24}\overline{C}^{(4)}_RH_4(\nu)+
\frac1{72}\left(\overline{C}^{(3)}_R\right)^2 H_6(\nu)
\nonumber\\
&&+ \frac{1}{120}\overline{C}^{(5)}_RH_5(\nu) 
+ \frac1{144}\overline{C}^{(3)}_R\overline{C}^{(4)}_R H_7(\nu) 
+ \frac1{1296}\left(\overline{C}^{(3)}_R\right)^3H_9(\nu)
+\ldots,\label{eq:Edgeworth}
\ee
%----------
where $H_n(\nu)$ is the $n$-th order Hermite polynomial, and we introduce normalized cumulants, 
$\overline{C}^{(n)}_R\equiv \bar\kappa^{(n)}_R/\bar\sigma_R^n$.
Since we are interested in the primordial non-Gaussianities arising from the coupling in the curvature perturbations 
higher than the quadratic order characterized by $\fnl$, 
we take account of the cumulants up to the fifth, $\bar\kappa_R^{(5)}$, such that we can discuss the effect from 
the quartic correction, $\hnl$.
We thus keep the series up to the fourth order in terms of the Edgeworth expansion (see Eq.~\ref{eq:Edgeworth}).
We follow the argument by \cite{LoVerde08} and
obtain the non-Gaussian halo mass function relevant to this order:
%----------
%\be
%\frac{\bar{f}(M,z)}{\bar{f}_{\rm G}(M,z)} &=& 
%\frac{{\rm PDF}_{\rm nG}(\nu)}{{\rm PDF}_{\rm G}(\nu)} 
%+\frac16\frac{{\rm d} (\bar\kappa^{(3)}_R/\sigma^3_R)}{{\rm d} \ln \sigma_R}\frac{1}{\nu}H_2(\nu)
%+\frac1{24}\frac{{\rm d} (\bar\kappa^{(4)}_R/\sigma^4_R)}{{\rm d} \ln \sigma_R}\frac{1}{\nu}H_3(\nu)\nonumber\\
%&&+\frac1{72}\frac{{\rm d} \bigl[(\bar\kappa^{(3)}_R)^2/\sigma^6_R\bigr]}{{\rm d} \ln \sigma_R}\frac{1}{\nu}H_5(\nu)
%+\frac1{120}\frac{{\rm d} (\bar\kappa^{(5)}_R/\sigma^5_R)}{{\rm d} \ln \sigma_R}\frac{1}{\nu}H_4(\nu)\nonumber\\
%&&+\frac1{144}\frac{{\rm d} (\bar\kappa^{(3)}_R\bar\kappa^{(4)}_R/\sigma^7_R)}{{\rm d} \ln \sigma_R}\frac{1}{\nu}H_6(\nu)
%+\frac1{1296}\frac{{\rm d} \bigl[(\bar\kappa^{(3)}_R)^3/\sigma^9_R\bigr]}{{\rm d} \ln \sigma_R}\frac{1}{\nu}H_8(\nu),
%\ee
%----------
%----------
\be
\frac{\bar{f}(M,z)}{\bar{f}_{\rm G}(M,z)} &=& 
\frac{{\rm PDF}_{\rm nG}(\nu)}{{\rm PDF}_{\rm G}(\nu)} 
+\frac16\frac{{\rm d}\,\overline{C}^{(3)}_R}{{\rm d} \ln \bar\sigma_R}\frac{1}{\nu}H_2(\nu)
+\frac1{24}\frac{{\rm d}\,\overline{C}^{(4)}_R}{{\rm d} \ln \bar\sigma_R}\frac{1}{\nu}H_3(\nu)\nonumber\\
&&+\frac1{72}\frac{{\rm d} (\overline{C}^{(3)}_R)^2}{{\rm d} \ln \bar\sigma_R}\frac{1}{\nu}H_5(\nu)
+\frac1{120}\frac{{\rm d}\,\overline{C}^{(5)}_R}{{\rm d} \ln \bar\sigma_R}\frac{1}{\nu}H_4(\nu)\nonumber\\
&&+\frac1{144}\frac{{\rm d} (\overline{C}^{(3)}_R\overline{C}^{(4)}_R)}{{\rm d} \ln \bar\sigma_R}\frac{1}{\nu}H_6(\nu)
+\frac1{1296}\frac{{\rm d} (\overline{C}^{(3)}_R)^3}{{\rm d} \ln \bar\sigma_R}\frac{1}{\nu}H_8(\nu),
\ee
%----------
where $\bar{f}_{\rm G}$ denotes the Gaussian mass function.
Note that as long as we consider the local-type non-Gaussianities, 
the derivative terms of $\overline{C}^{(n)}_R$ are usually very small given their weak 
$\bar\sigma_R$ dependence (see Eqs.~\ref{eq:fit_fnl}, \ref{eq:fit_gnl}, \ref{eq:fit_tnl}, 
and also Eqs.~\ref{eq:fit_hnl}, \ref{eq:fit_tnl2}, \ref{eq:fit_tnl3}), and the first term gives the dominant correction.
Note also that these derivative terms result in the correction terms in the bias parameters derived by 
\cite{Desjacques11a,Desjacques11b}.

Although the above ratio is derived based on the Press-Schechter formalism \cite{Press74}, 
we simply replace the Gaussian mass function with one recently calibrated by a large 
set of $N$-body simulations by \cite{Crocce10}:
%----------
\be
\bar{f}_{\rm G}(M,z) = f_{\rm MICE}(\bar\sigma_R,z) = A(z)\left[\bar\sigma_R^{-a(z)}+b(z)\right]\exp\left[-\frac{c(z)}
{\bar\sigma_R^2}\right],
\ee
%----------
where $A(z)=0.58(1+z)^{-0.13}$, $a(z)=1.37(1+z)^{-0.15}$, $b(z)=0.3(1+z)^{-0.084}$ and $c(z)=1.036(1+z)^{-0.024}$.
This mass function is shown to agree with $N$-body simulations over five orders of magnitude in the halo mass.

Although the mass function we adopt in this paper to incorporate the effect of the primordial non-Gaussianities 
might be too simplistic, one can easily refine the final result by replacing the mass function 
to a more accurate one.
See e.g., \cite{Matarrese00,Grossi09,Lam09,Maggiore10,Pillepich10,Valageas10,Desjacques10a} for various
elaborated halo mass functions for non-Gaussian initial conditions.

\bibliographystyle{JHEP} % style aa.bst
\bibliography{nGref,LSSref}

\providecommand{\href}[2]{#2}\begingroup\raggedright\begin{thebibliography}{10}

\bibitem{Bartolo04}
N.~{Bartolo}, E.~{Komatsu}, S.~{Matarrese}, and A.~{Riotto}, {\it
  {Non-Gaussianity from inflation: theory and observations}},  {\em Phys. Rep.}
  {\bf 402} (Nov., 2004) 103--266,
  [\href{http://xxx.lanl.gov/abs/astro-ph/}{{\tt astro-ph/}}].

\bibitem{Komatsu01}
E.~{Komatsu} and D.~N. {Spergel}, {\it {Acoustic signatures in the primary
  microwave background bispectrum}},  {\em Physical Review D} {\bf 63} (Mar.,
  2001) 063002, [\href{http://xxx.lanl.gov/abs/astro-ph/}{{\tt astro-ph/}}].

\bibitem{WMAP7}
E.~{Komatsu}, K.~M. {Smith}, J.~{Dunkley}, C.~L. {Bennett}, B.~{Gold},
  G.~{Hinshaw}, N.~{Jarosik}, D.~{Larson}, M.~R. {Nolta}, L.~{Page}, D.~N.
  {Spergel}, M.~{Halpern}, R.~S. {Hill}, A.~{Kogut}, M.~{Limon}, S.~S. {Meyer},
  N.~{Odegard}, G.~S. {Tucker}, J.~L. {Weiland}, E.~{Wollack}, and E.~L.
  {Wright}, {\it {Seven-year Wilkinson Microwave Anisotropy Probe (WMAP)
  Observations: Cosmological Interpretation}},  {\em Astrophys. J. S.} {\bf
  192} (Feb., 2011) 18, [\href{http://xxx.lanl.gov/abs/1001.4538}{{\tt
  arXiv:1001.4538}}].

\bibitem{Planck}
{The Planck Collaboration}, {\it {The Scientific Programme of Planck}},  {\em
  ArXiv Astrophysics e-prints} (Apr., 2006)
  [\href{http://xxx.lanl.gov/abs/astro-ph/}{{\tt astro-ph/}}].

\bibitem{Okamoto02}
T.~{Okamoto} and W.~{Hu}, {\it {Angular trispectra of CMB temperature and
  polarization}},  {\em Physical Review D} {\bf 66} (Sept., 2002) 063008,
  [\href{http://xxx.lanl.gov/abs/astro-ph/}{{\tt astro-ph/}}].

\bibitem{Enqvist05}
K.~{Enqvist} and S.~{Nurmi}, {\it {Non-Gaussianity in curvaton models with
  nearly quadratic potentials}},  {\em JCAP} {\bf 10} (Oct., 2005) 13,
  [\href{http://xxx.lanl.gov/abs/astro-ph/}{{\tt astro-ph/}}].

\bibitem{Boubekeur06}
L.~{Boubekeur} and D.~H. {Lyth}, {\it {Detecting a small perturbation through
  its non-Gaussianity}},  {\em Physical Review D} {\bf 73} (Jan., 2006) 021301,
  [\href{http://xxx.lanl.gov/abs/astro-ph/}{{\tt astro-ph/}}].

\bibitem{Suyama08}
T.~{Suyama} and M.~{Yamaguchi}, {\it {Non-Gaussianity in the modulated
  reheating scenario}},  {\em Physical Review D} {\bf 77} (Jan., 2008) 023505,
  [\href{http://xxx.lanl.gov/abs/0709.2545}{{\tt arXiv:0709.2545}}].

\bibitem{Suyama10}
T.~{Suyama}, T.~{Takahashi}, M.~{Yamaguchi}, and S.~{Yokoyama}, {\it {On
  classification of models of large local-type non-Gaussianity}},  {\em JCAP}
  {\bf 12} (Dec., 2010) 30, [\href{http://xxx.lanl.gov/abs/1009.1979}{{\tt
  arXiv:1009.1979}}].

\bibitem{Sugiyama11}
N.~S. {Sugiyama}, E.~{Komatsu}, and T.~{Futamase}, {\it {Non-Gaussianity
  Consistency Relation for Multifield Inflation}},  {\em Physical Review
  Letters} {\bf 106} (June, 2011) 251301,
  [\href{http://xxx.lanl.gov/abs/1101.3636}{{\tt arXiv:1101.3636}}].

\bibitem{Smith11c}
K.~M. {Smith}, M.~{Loverde}, and M.~{Zaldarriaga}, {\it {Universal Bound on
  N-Point Correlations from Inflation}},  {\em Physical Review Letters} {\bf
  107} (Nov., 2011) 191301, [\href{http://xxx.lanl.gov/abs/1108.1805}{{\tt
  arXiv:1108.1805}}].

\bibitem{Smidt10}
J.~{Smidt}, A.~{Amblard}, C.~T. {Byrnes}, A.~{Cooray}, A.~{Heavens}, and
  D.~{Munshi}, {\it {CMB contraints on primordial non-Gaussianity from the
  bispectrum (f$_{NL}$) and trispectrum (g$_{NL}$ and {$\tau$}$_{NL}$) and a
  new consistency test of single-field inflation}},  {\em Physical Review D}
  {\bf 81} (June, 2010) 123007, [\href{http://xxx.lanl.gov/abs/1004.1409}{{\tt
  arXiv:1004.1409}}].

\bibitem{Byrnes10a}
C.~T. {Byrnes}, S.~{Nurmi}, G.~{Tasinato}, and D.~{Wands}, {\it {Scale
  dependence of local f$_{NL}$}},  {\em JCAP} {\bf 2} (Feb., 2010) 34,
  [\href{http://xxx.lanl.gov/abs/0911.2780}{{\tt arXiv:0911.2780}}].

\bibitem{Byrnes10b}
C.~T. {Byrnes}, M.~{Gerstenlauer}, S.~{Nurmi}, G.~{Tasinato}, and D.~{Wands},
  {\it {Scale-dependent non-Gaussianity probes inflationary physics}},  {\em
  JCAP} {\bf 10} (Oct., 2010) 4, [\href{http://xxx.lanl.gov/abs/1007.4277}{{\tt
  arXiv:1007.4277}}].

\bibitem{Bernardeau02}
F.~{Bernardeau}, S.~{Colombi}, E.~{Gazta{\~n}aga}, and R.~{Scoccimarro}, {\it
  {Large-scale structure of the Universe and cosmological perturbation
  theory}},  {\em Phys. Rep.} {\bf 367} (Sept., 2002) 1--248,
  [\href{http://xxx.lanl.gov/abs/astro-ph/}{{\tt astro-ph/}}].

\bibitem{Kaiser87}
N.~{Kaiser}, {\it {Clustering in real space and in redshift space}},  {\em Mon.
  Not. Roy. Astron. Soc.} {\bf 227} (July, 1987) 1--21.

\bibitem{Kaiser84}
N.~{Kaiser}, {\it {On the spatial correlations of Abell clusters}},  {\em
  Astrophys. J. L.} {\bf 284} (Sept., 1984) L9--L12.

\bibitem{Scoccimarro00}
R.~{Scoccimarro}, {\it {Gravitational Clustering from {$\chi$}$^{2}$ Initial
  Conditions}},  {\em Astrophys. J.} {\bf 542} (Oct., 2000) 1--8,
  [\href{http://xxx.lanl.gov/abs/astro-ph/}{{\tt astro-ph/}}].

\bibitem{Verde00}
L.~{Verde}, L.~{Wang}, A.~F. {Heavens}, and M.~{Kamionkowski}, {\it
  {Large-scale structure, the cosmic microwave background and primordial
  non-Gaussianity}},  {\em Mon. Not. Roy. Astron. Soc.} {\bf 313} (Mar., 2000)
  141--147, [\href{http://xxx.lanl.gov/abs/astro-ph/}{{\tt astro-ph/}}].

\bibitem{Verde01}
L.~{Verde} and A.~F. {Heavens}, {\it {On the Trispectrum as a Gaussian Test for
  Cosmology}},  {\em Astrophysical J.} {\bf 553} (May, 2001) 14--24,
  [\href{http://xxx.lanl.gov/abs/astro-ph/}{{\tt astro-ph/}}].

\bibitem{Scoccimarro04b}
R.~{Scoccimarro}, E.~{Sefusatti}, and M.~{Zaldarriaga}, {\it {Probing
  primordial non-Gaussianity with large-scale structure}},  {\em Physical
  Review D} {\bf 69} (May, 2004) 103513,
  [\href{http://xxx.lanl.gov/abs/astro-ph/}{{\tt astro-ph/}}].

\bibitem{Sefusatti07}
E.~{Sefusatti} and E.~{Komatsu}, {\it {Bispectrum of galaxies from
  high-redshift galaxy surveys: Primordial non-Gaussianity and nonlinear galaxy
  bias}},  {\em Physical Review D} {\bf 76} (Oct., 2007) 083004,
  [\href{http://xxx.lanl.gov/abs/0705.0343}{{\tt arXiv:0705.0343}}].

\bibitem{Dalal08}
N.~{Dalal}, O.~{Dor{\'e}}, D.~{Huterer}, and A.~{Shirokov}, {\it {Imprints of
  primordial non-Gaussianities on large-scale structure: Scale-dependent bias
  and abundance of virialized objects}},  {\em Physical Review D} {\bf 77}
  (June, 2008) 123514, [\href{http://xxx.lanl.gov/abs/0710.4560}{{\tt
  arXiv:0710.4560}}].

\bibitem{Matarrese08}
S.~{Matarrese} and L.~{Verde}, {\it {The Effect of Primordial Non-Gaussianity
  on Halo Bias}},  {\em Astrophys. J.} {\bf 677} (Apr., 2008) L77--L80,
  [\href{http://xxx.lanl.gov/abs/0801.4826}{{\tt arXiv:0801.4826}}].

\bibitem{Afshordi08}
N.~{Afshordi} and A.~J. {Tolley}, {\it {Primordial non-Gaussianity, statistics
  of collapsed objects, and the integrated Sachs-Wolfe effect}},  {\em Physical
  Review D} {\bf 78} (Dec., 2008) 123507,
  [\href{http://xxx.lanl.gov/abs/0806.1046}{{\tt arXiv:0806.1046}}].

\bibitem{Taruya08b}
A.~{Taruya}, K.~{Koyama}, and T.~{Matsubara}, {\it {Signature of primordial
  non-Gaussianity on the matter power spectrum}},  {\em Physical Review D} {\bf
  78} (Dec., 2008) 123534, [\href{http://xxx.lanl.gov/abs/0808.4085}{{\tt
  arXiv:0808.4085}}].

\bibitem{McDonald08}
P.~{McDonald}, {\it {Primordial non-Gaussianity: Large-scale structure
  signature in the perturbative bias model}},  {\em Physical Review D} {\bf 78}
  (Dec., 2008) 123519, [\href{http://xxx.lanl.gov/abs/0806.1061}{{\tt
  arXiv:0806.1061}}].

\bibitem{Giannantonio10}
T.~{Giannantonio} and C.~{Porciani}, {\it {Structure formation from
  non-Gaussian initial conditions: Multivariate biasing, statistics, and
  comparison with N-body simulations}},  {\em Physical Review D} {\bf 81}
  (Mar., 2010) 063530, [\href{http://xxx.lanl.gov/abs/0911.0017}{{\tt
  arXiv:0911.0017}}].

\bibitem{Valageas10}
P.~{Valageas}, {\it {Mass function and bias of dark matter halos for
  non-Gaussian initial conditions}},  {\em A \& A} {\bf 514} (May, 2010) A46,
  [\href{http://xxx.lanl.gov/abs/0906.1042}{{\tt arXiv:0906.1042}}].

\bibitem{Desjacques11a}
V.~{Desjacques}, D.~{Jeong}, and F.~{Schmidt}, {\it {Non-Gaussian Halo Bias
  Re-examined: Mass-dependent Amplitude from the Peak-Background Split and
  Thresholding}},  {\em Physical Review D} {\bf 84} (Sept., 2011) 063512,
  [\href{http://xxx.lanl.gov/abs/1105.3628}{{\tt arXiv:1105.3628}}].

\bibitem{Desjacques11b}
V.~{Desjacques}, D.~{Jeong}, and F.~{Schmidt}, {\it {Accurate predictions for
  the scale-dependent galaxy bias from primordial non-Gaussianity}},  {\em
  Physical Review D} {\bf 84} (Sept., 2011) 061301,
  [\href{http://xxx.lanl.gov/abs/1105.3476}{{\tt arXiv:1105.3476}}].

\bibitem{Grossi09}
M.~{Grossi}, L.~{Verde}, C.~{Carbone}, K.~{Dolag}, E.~{Branchini},
  F.~{Iannuzzi}, S.~{Matarrese}, and L.~{Moscardini}, {\it {Large-scale
  non-Gaussian mass function and halo bias: tests on N-body simulations}},
  {\em Mon. Not. Roy. Astron. Soc.} {\bf 398} (Sept., 2009) 321--332,
  [\href{http://xxx.lanl.gov/abs/0902.2013}{{\tt arXiv:0902.2013}}].

\bibitem{Pillepich10}
A.~{Pillepich}, C.~{Porciani}, and O.~{Hahn}, {\it {Halo mass function and
  scale-dependent bias from N-body simulations with non-Gaussian initial
  conditions}},  {\em Mon. Not. Roy. Astron. Soc.} {\bf 402} (2010) 191--206,
  [\href{http://xxx.lanl.gov/abs/0811.4176}{{\tt arXiv:0811.4176}}].

\bibitem{Nishimichi10}
T.~{Nishimichi}, A.~{Taruya}, K.~{Koyama}, and C.~{Sabiu}, {\it {Scale
  dependence of halo bispectrum from non-Gaussian initial conditions in
  cosmological N-body simulations}},  {\em JCAP} {\bf 7} (July, 2010) 2,
  [\href{http://xxx.lanl.gov/abs/0911.4768}{{\tt arXiv:0911.4768}}].

\bibitem{Slosar08}
A.~{Slosar}, C.~{Hirata}, U.~{Seljak}, S.~{Ho}, and N.~{Padmanabhan}, {\it
  {Constraints on local primordial non-Gaussianity from large scale
  structure}},  {\em JCAP} {\bf 8} (Aug., 2008) 31,
  [\href{http://xxx.lanl.gov/abs/0805.3580}{{\tt arXiv:0805.3580}}].

\bibitem{Xia10a}
J.-Q. {Xia}, M.~{Viel}, C.~{Baccigalupi}, G.~{De Zotti}, S.~{Matarrese}, and
  L.~{Verde}, {\it {Primordial Non-Gaussianity and the NRAO VLA Sky Survey}},
  {\em Astrophys. J. L.} {\bf 717} (July, 2010) L17--L21,
  [\href{http://xxx.lanl.gov/abs/1003.3451}{{\tt arXiv:1003.3451}}].

\bibitem{Xia10b}
J.-Q. {Xia}, A.~{Bonaldi}, C.~{Baccigalupi}, G.~{De Zotti}, S.~{Matarrese},
  L.~{Verde}, and M.~{Viel}, {\it {Constraining primordial non-Gaussianity with
  high-redshift probes}},  {\em JCAP} {\bf 8} (Aug., 2010) 13,
  [\href{http://xxx.lanl.gov/abs/1007.1969}{{\tt arXiv:1007.1969}}].

\bibitem{Jeong09}
D.~{Jeong} and E.~{Komatsu}, {\it {Primordial Non-Gaussianity, Scale-dependent
  Bias, and the Bispectrum of Galaxies}},  {\em Astrophys. J.} {\bf 703} (Oct.,
  2009) 1230--1248, [\href{http://xxx.lanl.gov/abs/0904.0497}{{\tt
  arXiv:0904.0497}}].

\bibitem{Sefusatti09}
E.~{Sefusatti}, {\it {One-loop perturbative corrections to the matter and
  galaxy bispectrum with non-Gaussian initial conditions}},  {\em Physical
  Review D} {\bf 80} (Dec., 2009) 123002,
  [\href{http://xxx.lanl.gov/abs/0905.0717}{{\tt arXiv:0905.0717}}].

\bibitem{Baldauf11}
T.~{Baldauf}, U.~{Seljak}, and L.~{Senatore}, {\it {Primordial non-Gaussianity
  in the bispectrum of the halo density field}},  {\em JCAP} {\bf 4} (Apr.,
  2011) 6, [\href{http://xxx.lanl.gov/abs/1011.1513}{{\tt arXiv:1011.1513}}].

\bibitem{Sefusatti11}
E.~{Sefusatti}, M.~{Crocce}, and V.~{Desjacques}, {\it {The Halo Bispectrum in
  N-body Simulations with non-Gaussian Initial Conditions}},  {\em ArXiv
  e-prints} (Nov., 2011) [\href{http://xxx.lanl.gov/abs/1111.6966}{{\tt
  arXiv:1111.6966}}].

\bibitem{Desjacques10a}
V.~{Desjacques} and U.~{Seljak}, {\it {Signature of primordial non-Gaussianity
  of ${\phi_v}^{3}$ type in the mass function and bias of dark matter haloes}},
   {\em Physical Review D} {\bf 81} (Jan., 2010) 023006,
  [\href{http://xxx.lanl.gov/abs/0907.2257}{{\tt arXiv:0907.2257}}].

\bibitem{Smith11b}
K.~M. {Smith}, S.~{Ferraro}, and M.~{LoVerde}, {\it {Halo clustering and
  g\_$\{$NL$\}$-type primordial non-Gaussianity}},  {\em ArXiv e-prints} (June,
  2011) [\href{http://xxx.lanl.gov/abs/1106.0503}{{\tt arXiv:1106.0503}}].

\bibitem{Chongchitnan11}
S.~{Chongchitnan} and J.~{Silk}, {\it {Scale-dependent bias from the
  reconstruction of non-Gaussian distributions}},  {\em Physical Review D} {\bf
  83} (Apr., 2011) 083504, [\href{http://xxx.lanl.gov/abs/1012.1859}{{\tt
  arXiv:1012.1859}}].

\bibitem{Gong11}
J.-O. {Gong} and S.~{Yokoyama}, {\it {Scale-dependent bias from primordial
  non-Gaussianity with trispectrum}},  {\em Mon. Not. Roy. Astron. Soc.} {\bf
  417} (Oct., 2011) L79--L82, [\href{http://xxx.lanl.gov/abs/1106.4404}{{\tt
  arXiv:1106.4404}}].

\bibitem{Tseliakhovich10}
D.~{Tseliakhovich}, C.~{Hirata}, and A.~{Slosar}, {\it {Non-Gaussianity and
  large-scale structure in a two-field inflationary model}},  {\em Physical
  Review D} {\bf 82} (Aug., 2010) 043531,
  [\href{http://xxx.lanl.gov/abs/1004.3302}{{\tt arXiv:1004.3302}}].

\bibitem{Smith11a}
K.~M. {Smith} and M.~{LoVerde}, {\it {Local stochastic non-Gaussianity and
  N-body simulations}},  {\em JCAP} {\bf 11} (Nov., 2011) 9,
  [\href{http://xxx.lanl.gov/abs/1010.0055}{{\tt arXiv:1010.0055}}].

\bibitem{Becker11}
A.~{Becker}, D.~{Huterer}, and K.~{Kadota}, {\it {Scale-dependent
  non-Gaussianity as a generalization of the local model}},  {\em JCAP} {\bf 1}
  (Jan., 2011) 6, [\href{http://xxx.lanl.gov/abs/1009.4189}{{\tt
  arXiv:1009.4189}}].

\bibitem{Shandera11}
S.~{Shandera}, N.~{Dalal}, and D.~{Huterer}, {\it {A generalized local ansatz
  and its effect on halo bias}},  {\em JCAP} {\bf 3} (Mar., 2011) 17,
  [\href{http://xxx.lanl.gov/abs/1010.3722}{{\tt arXiv:1010.3722}}].

\bibitem{Yokoyama11b}
S.~{Yokoyama}, {\it {Scale-dependent bias from the primordial non-Gaussianity
  with a Gaussian-squared field}},  {\em JCAP} {\bf 11} (Nov., 2011) 1,
  [\href{http://xxx.lanl.gov/abs/1108.5569}{{\tt arXiv:1108.5569}}].

\bibitem{Schmidt10}
F.~{Schmidt} and M.~{Kamionkowski}, {\it {Halo clustering with nonlocal
  non-Gaussianity}},  {\em Physical Review D} {\bf 82} (Nov., 2010) 103002,
  [\href{http://xxx.lanl.gov/abs/1008.0638}{{\tt arXiv:1008.0638}}].

\bibitem{Wagner11}
C.~{Wagner} and L.~{Verde}, {\it {N-body simulations with generic non-Gaussian
  initial conditions II: Halo bias}},  {\em ArXiv e-prints} (Feb., 2011)
  [\href{http://xxx.lanl.gov/abs/1102.3229}{{\tt arXiv:1102.3229}}].

\bibitem{Scoccimarro11}
R.~{Scoccimarro}, L.~{Hui}, M.~{Manera}, and K.~C. {Chan}, {\it {Large-scale
  Bias and Efficient Generation of Initial Conditions for Non-Local Primordial
  Non-Gaussianity}},  {\em ArXiv e-prints} (Aug., 2011)
  [\href{http://xxx.lanl.gov/abs/1108.5512}{{\tt arXiv:1108.5512}}].

\bibitem{CAMB}
A.~{Lewis}, A.~{Challinor}, and A.~{Lasenby}, {\it {Efficient Computation of
  Cosmic Microwave Background Anisotropies in Closed Friedmann-Robertson-Walker
  Models}},  {\em Astrophys. J.} {\bf 538} (2000) 473--476,
  [\href{http://xxx.lanl.gov/abs/astro-ph/9911177}{{\tt astro-ph/9911177}}].

\bibitem{Chongchitnan10}
S.~{Chongchitnan} and J.~{Silk}, {\it {A Study of High-order Non-Gaussianity
  with Applications to Massive Clusters and Large Voids}},  {\em Astrophysical
  J.} {\bf 724} (Nov., 2010) 285--295,
  [\href{http://xxx.lanl.gov/abs/1007.1230}{{\tt arXiv:1007.1230}}].

\bibitem{LoVerde11}
M.~{LoVerde} and K.~M. {Smith}, {\it {The non-Gaussian halo mass function with
  f$_{NL}$, g$_{NL}$ and {$\tau$}$_{NL}$}},  {\em JCAP} {\bf 8} (Aug., 2011) 3,
  [\href{http://xxx.lanl.gov/abs/1102.1439}{{\tt arXiv:1102.1439}}].

\bibitem{Bardeen86}
J.~M. {Bardeen}, J.~R. {Bond}, N.~{Kaiser}, and A.~S. {Szalay}, {\it {The
  statistics of peaks of Gaussian random fields}},  {\em Astrophysical J.} {\bf
  304} (May, 1986) 15--61.

\bibitem{Cole89}
S.~{Cole} and N.~{Kaiser}, {\it {Biased clustering in the cold dark matter
  cosmogony}},  {\em Mon. Not. Roy. Astron. Soc.} {\bf 237} (Apr., 1989)
  1127--1146.

\bibitem{Mo96}
H.~J. {Mo} and S.~D.~M. {White}, {\it {An analytic model for the spatial
  clustering of dark matter haloes}},  {\em Mon. Not. Roy. Astron. Soc.} {\bf
  282} (Sept., 1996) 347--361, [\href{http://xxx.lanl.gov/abs/astro-ph/}{{\tt
  astro-ph/}}].

\bibitem{Catelan98}
P.~{Catelan}, F.~{Lucchin}, S.~{Matarrese}, and C.~{Porciani}, {\it {The bias
  field of dark matter haloes}},  {\em Mon. Not. Roy. Astron. Soc.} {\bf 297}
  (July, 1998) 692--712, [\href{http://xxx.lanl.gov/abs/astro-ph/}{{\tt
  astro-ph/}}].

\bibitem{Jenkins01}
A.~{Jenkins}, C.~S. {Frenk}, S.~D.~M. {White}, J.~M. {Colberg}, S.~{Cole},
  A.~E. {Evrard}, H.~M.~P. {Couchman}, and N.~{Yoshida}, {\it {The mass
  function of dark matter haloes}},  {\em Mon. Not. Roy. Astron. Soc.} {\bf
  321} (Feb., 2001) 372--384, [\href{http://xxx.lanl.gov/abs/astro-ph/}{{\tt
  astro-ph/}}].

\bibitem{Press74}
W.~H. {Press} and P.~{Schechter}, {\it {Formation of Galaxies and Clusters of
  Galaxies by Self-Similar Gravitational Condensation}},  {\em Astrophysical
  J.} {\bf 187} (Feb., 1974) 425--438.

\bibitem{Feldman94}
H.~A. {Feldman}, N.~{Kaiser}, and J.~A. {Peacock}, {\it {Power-spectrum
  analysis of three-dimensional redshift surveys}},  {\em Astrophys. J.} {\bf
  426} (1994) 23--37, [\href{http://xxx.lanl.gov/abs/astro-ph/9304022}{{\tt
  astro-ph/9304022}}].

\bibitem{EUCLID}
http://www.ias.u psud.fr/imEuclid/.

\bibitem{Sheth02}
R.~K. {Sheth} and G.~{Tormen}, {\it {An excursion set model of hierarchical
  clustering: ellipsoidal collapse and the moving barrier}},  {\em Mon. Not.
  Roy. Astron. Soc.} {\bf 329} (Jan., 2002) 61--75,
  [\href{http://xxx.lanl.gov/abs/astro-ph/}{{\tt astro-ph/}}].

\bibitem{Reid10}
B.~A. {Reid}, L.~{Verde}, K.~{Dolag}, S.~{Matarrese}, and L.~{Moscardini}, {\it
  {Non-Gaussian halo assembly bias}},  {\em JCAP} {\bf 7} (July, 2010) 13,
  [\href{http://xxx.lanl.gov/abs/1004.1637}{{\tt arXiv:1004.1637}}].

\bibitem{LoVerde08}
M.~{Lo Verde}, A.~{Miller}, S.~{Shandera}, and L.~{Verde}, {\it {Effects of
  scale-dependent non-Gaussianity on cosmological structures}},  {\em JCAP}
  {\bf 4} (Apr., 2008) 14, [\href{http://xxx.lanl.gov/abs/0711.4126}{{\tt
  arXiv:0711.4126}}].

\bibitem{Crocce10}
M.~{Crocce}, P.~{Fosalba}, F.~J. {Castander}, and E.~{Gazta{\~n}aga}, {\it
  {Simulating the Universe with MICE: the abundance of massive clusters}},
  {\em Mon. Not. Roy. Astron. Soc.} {\bf 403} (Apr., 2010) 1353--1367,
  [\href{http://xxx.lanl.gov/abs/0907.0019}{{\tt arXiv:0907.0019}}].

\bibitem{Matarrese00}
S.~{Matarrese}, L.~{Verde}, and R.~{Jimenez}, {\it {The Abundance of
  High-Redshift Objects as a Probe of Non-Gaussian Initial Conditions}},  {\em
  Astrophys. J.} {\bf 541} (Sept., 2000) 10--24,
  [\href{http://xxx.lanl.gov/abs/astro-ph/}{{\tt astro-ph/}}].

\bibitem{Lam09}
T.~Y. {Lam} and R.~K. {Sheth}, {\it {Halo abundances in the f$_{nl}$ model}},
  {\em Mon. Not. Roy. Astron. Soc.} {\bf 398} (Oct., 2009) 2143--2151,
  [\href{http://xxx.lanl.gov/abs/0905.1702}{{\tt arXiv:0905.1702}}].

\bibitem{Maggiore10}
M.~{Maggiore} and A.~{Riotto}, {\it {The Halo Mass Function from Excursion Set
  Theory. III. Non-Gaussian Fluctuations}},  {\em Astrophys. J.} {\bf 717}
  (July, 2010) 526--541, [\href{http://xxx.lanl.gov/abs/0903.1251}{{\tt
  arXiv:0903.1251}}].

\end{thebibliography}\endgroup

\end{document}